\documentclass[a4paper,11pt]{article}
\pdfoutput=1 

\usepackage{jcappub} 

\usepackage[T1]{fontenc} 
\usepackage{mathrsfs,fixmath} 
\usepackage{color}
\usepackage{graphicx}
\usepackage{subcaption}
\usepackage{amsmath}
\usepackage{breqn}
\usepackage{soul}

\def\Htrue{\mathscr{H}}
\def\gammaTrue{\gamma_{\mathscr{P}_A}}

\title{Component separation map-making for stochastic gravitational wave background}

\author[a]{Abhishek Parida,\note{Corresponding author.}}
\author[b]{Jishnu Suresh,}
\author[b]{Sanjit Mitra,}
\author[c]{Sanjay Jhingan}

\affiliation[a]{Centre for Theoretical Physics, Jamia Millia Islamia, New Delhi 110025, India}
\affiliation[b]{IUCAA, P. O. Bag 4, Ganeshkhind, Pune 411007, India}
\affiliation[c]{iCLA, Yamanashi Gakuin University, Japan}

\emailAdd{abhishek@ctp-jamia.res.in}
\emailAdd{jishnu@icrr.u-tokyo.ac.jp}
\emailAdd{sanjit@iucaa.in}
\emailAdd{sanjay.jhingan@gmail.com}

\abstract{
Superposition of unresolved sources of gravitational waves (GW) is expected to create a persistent background of stochastic gravitational wave background (SGWB). Different types of astrophysical and cosmological sources are however likely to be present in the background. We present an algorithm for separation of the components with distinct frequency spectral indices into individual sky-maps. To demonstrate the method, we simulate GW signals for different spectral indices and corresponding sky-maps, e.g., point sources like the Virgo cluster and extended ones like the Milky Way Galaxy, and apply our method to recover the injected maps.}

\begin{document}
\maketitle
\flushbottom

\section{Introduction}

Gravitational Wave (GW) astronomy had an exciting beginning. In the first two years, ten binary black hole mergers \cite{GWTC1} and one binary neutron star merger have been observed~\cite{GW170817}; the latter was followed up several electromagnetic telescopes across the world in multiple wavelengths, marking the beginning of multi-messenger astronomy with GW. Many kinds of sources are however expected to be detected with GW observations, Stochastic Gravitational Wave Background (SGWB) \cite{1,2,3,4} being one of the most interesting ones. An SGWB is expected to arise from unresolved Astrophysical sources and processes occurring in the early Universe \cite{reg1,reg2}. The binary merger rates estimated from the observed signals indicate that the isotropic SGWB created by compact binary coalescence may be detected by present ground-based interferometric detectors in few years of observation.

An astrophysical background can be significantly anisotropic if it is dominated by the local universe~\cite{n_mazumder}. Aperture synthesis imaging is a general approach often used to produce images with high angular resolution in the field of astronomy and bears a generic name of \textit{Radiometry} in literature. It uses the fundamental idea of interference of plane waves to locate objects on the celestial sphere. In the context of gravitational radiation, the signal from a patch (a specific direction) on the sky, arriving at two or more detector sites with a relative phase is cross-correlated with a corresponding relative time shift in the process of map making. Due to earth's rotation, the phase differences (corresponding to the time-shifts) vary with time \cite{radiometer_a, radiometer_b, radiometer_c, radiometer_d, radiometer_e}.
%
%
GW radiometer algorithm can be implemented in either pixel or spherical harmonic basis and are routinely applied to LIGO data \cite{radiometer_data_a,radiometer_data_b,radiometer_data_c,radiometer_data_d,radiometer_data_e}.

The component separation problem, in general, can be stated as: given many observations of the sky, how one can isolate the contribution, in the total observed emission, of all the different astrophysical processes, which in turn contribute to the total datasets. In signal processing language, this type of problem is typically treated on the basis of statistical tools, which assume the total contribution arises from the linear superposition of a number of independent components, or sources, as in the case of SGWB. Component separation is perceived to be one of the key challenges once we detect an SGWB in the future. While it is always possible to build more sensitive instruments to detect the background, at the end of the day, astrophysical confusion will be the primary source of uncertainty that exists in the discovery. This requires developing the data analysis methods needed to address the issue optimally.

In the case of SGWB, the background consist of various types of Astrophysical and Cosmological sources which are modeled in the frequency spectra, considering the former as a dominant contributor. These models provide an observational limit on the free parameters involved in the theory. Mostly, the searches made on the LIGO data assumes a single frequency spectrum background. However, in an earlier work, it was demonstrated using toy models that, in the case of \textit{multiple} isotropic background, the measurements of the amplitude would always be overestimated. These amplitudes represented different components of the GW spectrum each with a known spectral shape. In our method, we propose to constrain these \textit{components} of a background jointly, by analytically transforming the Maximum Likelihood estimation problem to a linear deconvolution problem. In this work, we extend our approach from \textit{isotropic} to the \textit{directed search}.

In the past few years, the efficiency of GW radiometer algorithm has been improved dramatically by the introduction of data folding \cite{folding} as well as a very fast efficient new method called PyStoch \cite{pyStoch}. The temporal symmetry in the geometric part of the GW radiometry algebra has been utilized to fold the entire detector data of several hundreds of days to only one sidereal day (i.e., 23 hr 56 min 4 sec), thereby reducing the computational cost by a factor equal to the total number of days of observation. On the other hand, PyStoch considered the compactness of the folded data and replaced the loops in the pipeline with matrix multiplications. Also incorporated the popular HEALPix pixelization tools for further standardization and optimization. Folding and PyStoch together has made the entire stochastic gravitational wave background analysis a few thousand times faster. We incorporated the full advantage of these two methods in this work.

Organization of the paper is as follows. In section \ref{sub_sec:general_characteristic_of_signal}, we begin by characterizing the brightness of an extended source on the sky, and its relation to the spectrum of an SGWB. We present the essentials of GW Radiometry in section \ref{sub_sec:working_principle} particularly the standard methods of cross-correlation statistics for directional searches. Finally, we illustrate the general idea underlying our formalism of \textit{component separation}, to include more than one background in section \ref{sub_sec:component_separation}. We present results in section \ref{sec:results} and write a summary on utilizing the HEALPix and Pystoch module. Finally, we conclude with a discussion in section \ref{sec:conclusion}.

\section{Formalism}
\subsection{General characteristic of the signal from an anisotropic SGWB}\label{sub_sec:general_characteristic_of_signal}

Assuming the signal to be stochastic and uncorrelated in the two polarizations and different frequencies and directions, the two-point correlation function for the Fourier modes of the GW signal can be written as
\begin{equation}\label{eqn:fourier_modes}
 \langle \tilde{h}_A^*(f,\hat{\mathbf{\Omega}}) \ \tilde{h}_A(f',\hat{\mathbf{\Omega'}})  \rangle =  \delta_{AA'} \ \delta (f-f') \ \delta^2 (\mathbf{\hat{\Omega}} - \hat{\mathbf{\Omega'}}) \ \mathscr{I}_A(\mathbf{\hat{\Omega}}, f) \, ,
\end{equation}
where $\mathscr{I}_A(\hat{\mathbf{\Omega}}, f)$ is a quantity that is proportional to the intensity of a SGWB for a particular polarization, $A = [+, \times]$ and in general depends on frequency $f$ and direction of the source $\hat{\mathbf{\Omega}}$, in the sky.
%
%
The \textit{true} frequency power spectrum $\mathscr{H}(f)$ can be assumed to fairly remain constant and separable from its angular dependency as,
\begin{equation}
\mathscr{I}_A(\hat{\mathbf{\Omega}}, f) = \mathscr{H}(f) \, \mathscr{P}_A(\hat{\mathbf{\Omega}}) \,.
\end{equation}
$\mathscr{P}_A(\hat{\mathbf{\Omega}})$ is related to specific intensity of a SGWB, $I_{GW}(\hat{\mathbf{\Omega}}, f)$ \cite{radiometer_d}, and commonly called the ``source map''. 
\begin{equation}
I_{GW}(\mathbf{\hat{\Omega}}, f) = \frac{4 \pi^2 c}{3 H_0^2} \ f^2 \ \sum_A \ \mathscr{I}_A(\hat{\mathbf{\Omega}}, f),    
\end{equation}
where $c$ denotes the speed of light and $H_0$, is the value of Hubble parameter at the present epoch. Throughout the paper we denote the model spectrum as $H(f)$ and model angular power distribution as $\mathcal{P}(\hat{\mathbf{\Omega}})$, to distinguish from their corresponding \textit{true} values. From a detection point of view, the time series output $s_1(t)$ and $s_2(t)$ of two GW detectors is viewed as a combination of the celestial signal $h_1(t)$, $h_2(t)$ contaminated with terrestrial noise $n_1(t)$ and $n_2(t)$ respectively. The time series data is broken into small time units $\Delta T \sim$ few minutes and analysis is done with the Fourier transforms of these smaller segments, often called Short Fourier Transforms (SFT). The duration of each SFT units are much greater than the light travel time between the two detectors and at the same time it is smaller compared to the time over which the detector noise spectrum and the Earth can be regarded as stationary. Furthermore, noises at two detector sites are assumed to be uncorrelated with each other and with the GW strain signal.

\begin{equation}
s_I(t) = h_I(t) + n_I(t), \quad I = 1,2 
\end{equation}

\begin{equation}
s_I(t;f) = \int_{t-\Delta T/2}^{t+\Delta T/2} dt' \ \tilde{s}_I(t') \ e^{-i2\pi f t'} = \tilde{h}_I(t;f) + \tilde{n}_I(t;f) \, ,
\end{equation}
\begin{equation}
\langle \tilde{h}_I^*(t;f) \ \tilde{n}_I(t;f)  \rangle = 0 =  \langle \tilde{n}_1^*(t;f) \ \tilde{n}_2(t;f)  \rangle \, .
\end{equation}

In the limit of small signal amplitude, the cross-correlation relation takes a simple approximate form as,  

\begin{equation}\label{eqn:csd}
\langle \tilde{s}_1^*(t;f) \ \tilde{s}_2(t';f')  \rangle \approx \langle \tilde{h}_1^*(t;f) \ \tilde{h}_2(t';f')  \rangle = \Delta T \ \delta_{tt'} \ \delta_{ff'} \ \mathscr{H}(f) \ \gammaTrue(t, f).
\end{equation}
we refer this combination (product) of the complex conjugate of SFTs from two detectors as Cross-power Spectral Density (CSD). As mentioned earlier, $\Htrue(f)$ represent the two sided Power Spectral Density (PSD) and $\gammaTrue(t,f)$ is overlap-reduction-function (orf) corresponding to the \textit{true} background,

\begin{subequations}\label{eqn:subeqns}
\begin{align}
\gammaTrue(t, f) &= \int_{S^2} d \hat{\mathbf{\Omega}} \ \bigg[\sum_{A = [+, \times]} F^A_1(\hat{\mathbf{\Omega}}, t) F^A_2(\hat{\mathbf{\Omega}}, t) \ \mathscr{P}^A(\hat{\mathbf{\Omega}}) \bigg] \cdot e^{2 \pi i f (\hat{\mathbf{\Omega}} \cdot \mathbf{\Delta x}(t)/c)} \label{eqn:subeq1}\\
&= \int_{S^2} d \hat{\mathbf{\Omega}} \ \Gamma(\hat{\mathbf{\Omega}}, t) \cdot \mathscr{P}(\hat{\mathbf{\Omega}}) \cdot e^{2 \pi i f (\hat{\mathbf{\Omega}} \cdot \mathbf{\Delta x}(t)/c)} \label{eqn:subeq2}\\
&\equiv  \int_{S^2} d \hat{\mathbf{\Omega}} \ \gamma(\hat{\mathbf{\Omega}}, t, f) \cdot \mathscr{P}(\hat{\mathbf{\Omega}})\label{eqn:subeq3}
\end{align}
\end{subequations}
where $\Gamma(\hat{\mathbf{\Omega}}, t) = \sum_{A = [+, \times]} F^A_1(\hat{\mathbf{\Omega}}, t) F^A_2(\hat{\mathbf{\Omega}}, t)$, is the combined antenna pattern function of the pair of baseline detectors. Each of these functions are defined as a scalar product between the polarization tensor and the detector tensor as $F^A_I(\hat{\mathbf{\Omega}}, t) := e^A_{ij}(\hat{\mathbf{\Omega}}) \ d^{ij}_I(t)$. In going from equation $\eqref{eqn:subeq1}$ to $\eqref{eqn:subeq2}$, we assumed equal power in both the polarization, i.e., $\mathscr{P}^+(\hat{\mathbf{\Omega}}) = \mathscr{P}^{\times}(\hat{\mathbf{\Omega}}) = \mathscr{P}(\hat{\mathbf{\Omega}})$ (say). Finally, the equation $\eqref{eqn:subeq3}$, is a convenience used to separate the information of the \textit{true} GW-sky, $\mathscr{P}(\hat{\mathbf{\Omega}})$, from the  antenna pattern function. The arguments in the orf, $\gamma(\hat{\mathbf{\Omega}}, t, f)$, denotes that it is a function of $t$, $f$ and the pixel index, $\hat{\mathbf{\Omega}} \equiv \hat{\mathbf{\Omega}}_i$, i.e., the corresponding direction in the sky and $\mathbf{\Delta} \mathbf{x} (t) := \mathbf{x}_2 (t) - \mathbf{x}_1 (t)$ is the baseline separation vector between the pair of detectors. It is worth mentioning the relation of the PSD to a more popular and dimensionless quantity $\Omega_{\text{GW}}(f)$, which is the spectrum of an isotropic SGWB . It is defined as the energy density ($\rho_{\text{GW}}$) per unit logarithmic frequency interval $f,f+df$. Using the definition of $\Omega_{\text{GW}}(f)$ one can obtain its relation to the PSD as  
\begin{equation}
\Omega_{\text{GW}}(f) \ = \ \frac{4\pi^2}{3H_0^2} \ f^3 \ H(f) \ \sum_A \ \int_{S^2} d\hat{\mathbf{\Omega}} \ \mathcal{P}_A(\hat{\mathbf{\Omega}}) \ = \ \frac{32\pi^3}{3H_0^2} \ f^3 \ H(f) .
\end{equation}
The last step is valid for an \textit{isotropic} background which is arrived at by utilizing the direction independence for the polarization fields. The integral yields a factor of $8 \pi$.

\subsection{Working principle of Gravitational Wave Radiometry}\label{sub_sec:working_principle}
\begin{figure}[h]
\centering
\includegraphics[width=0.6\textwidth]{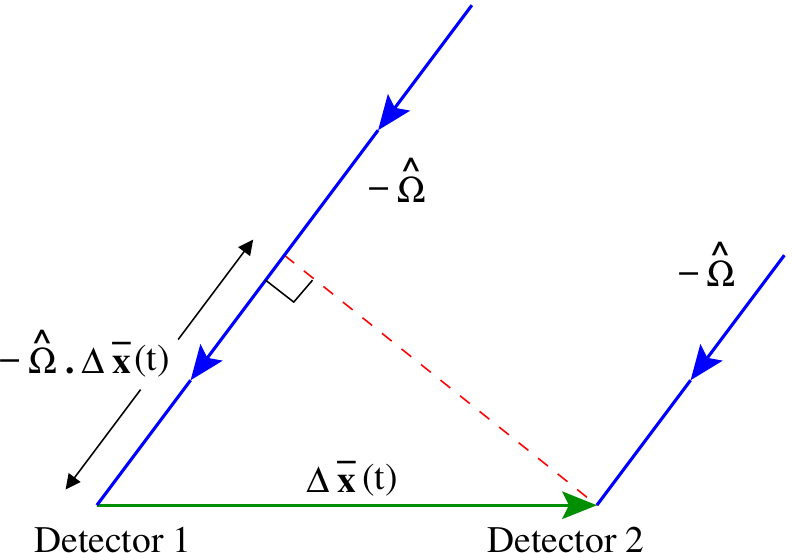}
\caption{Geometry of baseline: schematic diagram of the radiometer. To measure the signal coming from a direction $\mathbf{\hat{\Omega}}$, data from the detectors are correlated with a phase delay of $\mathbf{\hat{\Omega}}\cdot {\mathbf{\Delta x}_I (t)}/c$. \cite{radiometer_d}}
\label{fig:radiometry}
\end{figure}

Interferometers performing a radiometer analysis, separated by a baseline distance would measure the correlation between the time series signals collected by different antennas. Consider the arrival of several plane wavefront from a fixed source position in the sky onto geographically separated detector sites, each with a position vector \textbf{x}$_I$ where $I = 1,2$, refer figure (\ref{fig:radiometry}). These signals at any site encode a relative time difference due to the propagating plane waves from the source direction, which translates to a relative phase difference. This is the basic principle behind searching for anisotropy in a stochastic GWB. A similar algorithm is commonly used in radio astronomy and cosmic microwave background observations. Let $\hat{\mathbf{\Omega}}$ be the usual notation for the direction of the unit vector pointing to the source location in the celestial equatorial frame. In this coordinate system, the interferometers rotate along with the rotation of Earth, but the magnitude of the baseline $\mathbf{\Delta x}$ remains constant. Upon cross-correlating, the data from a pair of detectors, one being time-delayed to the other, would cause the signals arriving from a particular direction to interfere constructively \cite{radiometer_d}. This is the main idea behind capturing anisotropies and constructing a sky map. 

GW sky map is computed in two steps. Firstly, the cross-correlated signal from the detectors (CSDs) is multiplied with a suitable, directional dependent filter and integrated to obtain an intermediate statistic for every time segment. Then the final statistic, also known as the \textit{dirty map} $S(\hat{\mathbf{\Omega}})$, is attained by linearly combining the statistics over the smaller time intervals as a weighted sum.
%
%

%
\begin{equation}\label{eqn:estimate_map}
\langle S_{\alpha}(\hat{\mathbf{\Omega}}) \rangle = \Lambda_{\alpha}(\hat{\mathbf{\Omega}}) \times \bigg[2 \sum_{t f} \langle \tilde{s}_1^*(t;f) \ \tilde{s}_2(t;f) \rangle \bigg(\frac{H_{\alpha}(f) \ \gamma^*(\hat{\mathbf{\Omega}}, t, f)}{P_1(t; f) P_2(t; f)}\bigg)\bigg],
\end{equation}
where $\gamma(\hat{\mathbf{\Omega}}, t, f)$ is the direction dependent reduction function between the two detector sites, defined in equation $\eqref{eqn:subeq3}$, and the normalizing constant, $\Lambda_{\alpha}(\hat{\mathbf{\Omega}})$, a summation on time and frequency would appear several times in the article henceforth.

\begin{equation}
\Lambda_{\alpha}(\hat{\mathbf{\Omega}}) =  \bigg[\sum_{t f} \bigg( \frac{H^2_\alpha(f) \ |\gamma(\hat{\mathbf{\Omega}}, t, f)|^2}{P_1(t; f) P_2(t; f)}\bigg)\bigg]^{-1} 
\end{equation}
This equation is an estimate for the GW sky for a given spectral shape $H_\alpha(f)$. The index $\alpha$ is an identifier for the spectral shape that one is investigating. We refer to equation (\ref{eqn:estimate_map}) as the \textit{single-index} estimate, for the reasons to become clearer later. This sky-estimate can also be expressed as a convolution of the \textit{source map} with the \textit{beam pattern function}, along with added statistical noise. The \textit{beam}, expressed as a square matrix, is the response of the baseline detectors for a point source in the sky. The expression for the beam matrix is given by~\cite{radiometer_d}, 

\begin{equation}\label{eqn:beam}
B_{\alpha}(\hat{\mathbf{\Omega}}, \hat{\mathbf{\Omega}}') = \Lambda_{\alpha}(\hat{\mathbf{\Omega}}) \times \bigg[\sum_{tf} \bigg(\frac{\mathscr{H}(f) \ H_\alpha(f)}{P_1(t;f) \ P_2(t;f)}\bigg) \ \Gamma(\hat{\mathbf{\Omega}}', t) \ \Gamma(\hat{\mathbf{\Omega}}, t) \ e^{-2 \pi i f (\Delta \hat{\mathbf{\Omega}} \cdot \mathbf{\Delta x}(t)/c)}\bigg]
\end{equation}
The suffix $\alpha$ distinguishes among the various model spectrum and $\Delta \hat{\mathbf{\Omega}} = |\hat{\mathbf{\Omega}}-\hat{\mathbf{\Omega}}'|$. As remarked earlier, $\mathscr{H}(f)$ is the \textit{true} spectrum and $H_\alpha(f)$ denotes various model spectra. An account of few anticipated spectrum is profiled in section (\ref{sub_sec:component_separation}). The dirty map can then be expressed by the linear convolution equation,
\begin{equation}\label{eqn:conv}
S_{\alpha}(\hat{\mathbf{\Omega}}) = B_{\alpha}(\hat{\mathbf{\Omega}}, \hat{\mathbf{\Omega}}') \cdot \mathscr{P}_A(\hat{\mathbf{\Omega}}') + \text{statistical noise}, 
\end{equation}

Secondly, to estimate the true sky, one needs to reverse the effects of convolution on the recorded data using a deconvolution algorithm. One can estimate the Maximum Likelihood sky-map for the \textit{clean map} $\hat{\mathcal{P}}_A(\hat{\mathbf{\Omega}})$, by directly inverting the beam matrix. However, because the beam matrix is mostly sparse, one has to rely on other techniques for solving the linear convolution equation (\ref{eqn:conv}). Among several methods, namely, matrix inversion of the beam, least-square, minimum residual we affirmed that the conjugate gradient method was most successful in reliably reproducing an estimate for the underlying sky map.


\subsection{Extension to Component Separation}\label{sub_sec:component_separation}

In our previous work, we showed using the isotropic background that \textit{single-index} search scheme will always overestimate the $true$ values in the presence of a multi-component background. It was illustrated in detail with the help of toy models, employing two and three component background and utilizing initial and advance LIGO sensitivity \cite{compSep}. We injected signals corresponding to realistic values for the dimensionless parameter $\Omega_{\alpha}$ and attempted to recover them.
In the present work, we perform the same for directed search.

\subsubsection{Multiple component background}\label{sub_sec:multiComponents}
An SGWB comprises of Cosmological and Astrophysical sources, and a spectral index $\alpha$ is often used to distinguish among the various models. The method presented here is applicable for any general spectral shape as indicated by $H_{\alpha}(f)$, where $\alpha$ is attributed to backgrounds concerning different spectral shapes. However, for the numerical purposes, we make use of the power-law spectrum, $\mathcal{F}^{\alpha}(f) = \big(\frac{f}{f_{\text{ref}}}\big)^{\alpha}$, where $f_{\text{ref}} = 100$ Hz. Below are some anticipated power-law backgrounds, some popular \textit{components} found in the literature. We also provide in the table their corresponding model spectrum, shown as merely indicative form. It would be interesting to find out a different spectrum aside from those motivated by physics.

\begin{table}[h!]
\centering
\begin{tabular}{|c|c|c|}\hline\hline
Spectral index $\alpha$ & $H_{\alpha}(f) \sim \mathcal{F}^{\alpha}(f) \cdot f^{-3}$ & Representation \\\hline \hline
0 & $f^{-3}$ & slow-roll inflationary scenario \\\hline
2/3 & $f^{-7/3} \cdot f_{\text{ref}}^{-2/3}$ & coalescence of compact binary objects \\\hline
3 & $f_{\text{ref}}^{-3}$ & single rotating neutron stars \\\hline\hline
\end{tabular}
\caption {A depiction of some familiar Cosmological and Astrophysical phenomenon; column 2 refers to the functional form of the spectrum used for numerical analysis.}\label{tab:alpha_representation}
\end{table}

A component of GW sky, for a directional search, is often characterized by the combination of a specific spectral shape, i.e., PSD, $H_{\alpha}(f)$, and its corresponding angular dependency, i.e., how the source luminosity is distributed in the 2-D sky, $\mathcal{P}^{\alpha}(\hat{\mathbf{\Omega}})$. The intensity of SGWB can be comprised of various \textit{components} as follows\footnote{We remove the suffix $A$ representing the polarization due to the assumption of equal contribution in each polarization as discussed in section (\ref{sub_sec:general_characteristic_of_signal}).},
\begin{equation}\label{eqn:multi-background}
\mathscr{I}(\hat{\mathbf{\Omega}}, f) = \sum_{\alpha} H_{\alpha}(f) \cdot \mathcal{P}^{\alpha}(\hat{\mathbf{\Omega}})
\end{equation}
Now we are in a situation to present the linear convolution equation for the multiple-background case. A convolution equation often has viewed as an expression for a linear measurement phenomenon, where the signal from nature is integrated with the point-spread function, in our case the \textit{beam matrix}, of the telescope (or antenna) to yield an intermediate result. In the present work, the convolution is carried in the pixel domain as opposed to the usual Fourier domain often used in radio astronomy, and the intermediate result is the \textit{dirty map}. Moreover, examining the above equation (\ref{eqn:estimate_map}) and (\ref{eqn:conv}), one can easily reflect that for a single background case, i.e., a SGWB with only one \textit{component}, say indicated by the spectral index $\alpha = \alpha_1$, the statistics in equation (\ref{eqn:estimate_map}) yields the component-map precisely up to a constant factor.

\begin{subequations}
\begin{align}
\mathscr{H}(f) \cdot \mathscr{P}(\hat{\mathbf{\Omega}}) &= H_{\alpha_1}(f) \cdot \mathcal{P}_{\alpha_1}(\hat{\mathbf{\Omega}}) \label{eqn:condition}\\ 
\Rightarrow [S_{\alpha_1}(\hat{\mathbf{\Omega}})] &= [B_{\alpha_1}(\hat{\mathbf{\Omega}}, \hat{\mathbf{\Omega}}')] \ [\mathcal{P}_{\alpha_1}(\hat{\mathbf{\Omega}}')]\label{eqn:effect}\\
\text{and} \quad \mathcal{\hat{P}}_{\alpha_1}(\hat{\mathbf{\Omega}}) &= [B_{\alpha_1}(\hat{\mathbf{\Omega}}, \hat{\mathbf{\Omega}}')]^{-1} \ [S_{\alpha_1}(\hat{\mathbf{\Omega}}')].\label{eqn:output_map}
\end{align}
\end{subequations}
Equation (\ref{eqn:condition}) represents an optimal scenario when the combination of the PSD and source angular luminosity is proportional to the modeled background; the resulting SNR is maximum. Otherwise, it is sub-optimal.
The impact on the statistics, presented in equation (\ref{eqn:effect}) due to the condition in equation (\ref{eqn:condition}) was verified in numerical simulations and forms the basis for our study. In doing so, we injected a single point source (hot-spot) with a specific known spectral shape (power-law) and were able to recover the hot-spot, although there is smearing out effect as seen in figure (\ref{fig:comparison_Single_index}). Equation (\ref{eqn:output_map}) is a mere representation of the deconvolution operation to estimate the output, namely the \textit{clean-map} $\mathcal{\hat{P}}_{\alpha_1}(\hat{\mathbf{\Omega}})$. However, for an extended source, the effect is not straightforward as there will be contributions from nearby pixels entering in the exponential term through $\Delta \hat{\mathbf{\Omega}}$, refer to equation (\ref{eqn:beam}). 

\subsubsection{Towards a Coupling matrix}
Emulating our previous work on \textit{isotropic} backgrounds, we introduce an additional index $\alpha$, refer to equation (\ref{eqn:multi-background}), to incorporate correlation among various spectral shapes and motivate for the core convolution equation. We provide the necessary steps to modify the above equations for the case of multiple backgrounds. 
First, we decompose the CSD such that the contribution from various GW processes, distinguished by spectral shapes, (\textit{components}) are noticeable. Substituting equation (\ref{eqn:multi-background}) in equation (\ref{eqn:fourier_modes}), one can arrive at an expression for the CSD as follows,

\begin{eqnarray}\label{eqn:csd_decomposition}
\langle \tilde{h}_1^*(t;f) \ \tilde{h}_2(t;f) \rangle &=& \Delta T \ \sum_{\alpha} H_{\alpha}(f) \ \int_{S^2} d \hat{\mathbf{\Omega}} \ \gamma(\hat{\mathbf{\Omega}},f,t) \cdot \mathcal{P}^{\alpha}(\hat{\mathbf{\Omega}}) \nonumber \\
&=&\Delta T \int_{S^2} d \hat{\mathbf{\Omega}} \ \gamma(\hat{\mathbf{\Omega}}, t) \ \bigg\{H_{\alpha_1}(f) \cdot \mathcal{P}^{\alpha_1}(\hat{\mathbf{\Omega}}) + H_{\alpha_2}(f) \cdot \mathcal{P}^{\alpha_2}(\hat{\mathbf{\Omega}}) \nonumber \\
&& + H_{\alpha_3}(f) \cdot \mathcal{P}^{\alpha_3}(\hat{\mathbf{\Omega}}) + \cdot\cdot\cdot + H_{\alpha_n}(f) \cdot \mathcal{P}^{\alpha_n}(\hat{\mathbf{\Omega}})\bigg\}   
\end{eqnarray}    
We make use of the equation (\ref{eqn:csd_decomposition}) in equation (\ref{eqn:csd}) and present the algebra with n-components. The application of successive-distinct directional dependent filters with the spectral identifier, $\alpha = \{\alpha_1, \alpha_2, \alpha_3, \cdot \cdot \cdot, \alpha_n\}$, would provide estimates for the \textit{dirty-maps}, $\langle S_{\alpha_1}(\hat{\mathbf{\Omega}}) \rangle, \langle S_{\alpha_2}(\hat{\mathbf{\Omega}}) \rangle, \langle S_{\alpha_3}(\hat{\mathbf{\Omega}}) \rangle, \cdot \cdot \cdot, S_{\alpha_n}(\hat{\mathbf{\Omega}}) \rangle$, of the GW sky for a particular spectrum. As mentioned earlier these \textit{single-component} estimates overestimate the evaluation. The entire step can be summarized via the same convolution equation (\ref{eqn:conv}), but this time it is viewed as a combination of several equations, namely a matrix equation. Now we formulate the corresponding integral equation for a multi-background, by introducing a general prescription. Substituting the CSD decomposition, equation (\ref{eqn:csd_decomposition}), in equation (\ref{eqn:estimate_map}) and applying different filters progressively one arrives at the master equation. First, we present it for the single shape $\alpha = \alpha_1$. 

\begin{eqnarray}
\langle S_{\alpha_1}(\hat{\mathbf{\Omega}}_i) \rangle &=& \Lambda_{\alpha_1}(\hat{\mathbf{\Omega}}_i) \times \bigg[2 \ \Delta T \sum_{t f} \int_{S^2} d \hat{\mathbf{\Omega}}_j \ \gamma(\hat{\mathbf{\Omega}}_j, t) \bigg\{H_{\alpha_1}(f) \cdot \mathcal{P}^{\alpha_1}(\hat{\mathbf{\Omega}}_j) \nonumber \\
&& + H_{\alpha_2}(f) \cdot \mathcal{P}^{\alpha_2}(\hat{\mathbf{\Omega}}_j) + H_{\alpha_3}(f) \cdot \mathcal{P}^{\alpha_3}(\hat{\mathbf{\Omega}}_j) + \cdot\cdot\cdot \nonumber \\
&& + H_{\alpha_n}(f) \cdot \mathcal{P}^{\alpha_n}(\hat{\mathbf{\Omega}}_j)\bigg\} \bigg(\frac{H_{\alpha_1}(f) \ \gamma^*(\hat{\mathbf{\Omega}}_i, t, f)}{P_1(t; f) P_2(t; f)}\bigg)\bigg]
\end{eqnarray}

\begin{eqnarray}
\langle S_{\alpha_1}(\hat{\mathbf{\Omega}}_i) \rangle &=& \Lambda_{\alpha_1}(\hat{\mathbf{\Omega}}_i) \times 2 \ \Delta T \
\sum_{t f} \int_{S^2} d \hat{\mathbf{\Omega}}_j \ \frac{H_{\alpha_1}(f) \ \gamma^*(\hat{\mathbf{\Omega}}_i, t, f) \ \gamma(\hat{\mathbf{\Omega}}_j, t, f)}{P_1(t; f) P_2(t; f)} \nonumber \\
&& \times \begin{bmatrix}
H_{\alpha_1}(f) & H_{\alpha_2}(f) & H_{\alpha_3}(f) & \cdot \cdot \cdot & H_{\alpha_n}(f)
\end{bmatrix} \cdot 
\begin{bmatrix}
\mathcal{P}^{\alpha_1}(\hat{\mathbf{\Omega}}_j) \\ \mathcal{P}^{\alpha_2}(\hat{\mathbf{\Omega}}_j) \\ \mathcal{P}^{\alpha_3}(\hat{\mathbf{\Omega}}_j) \\ \cdot \\ \cdot \\ \cdot \\ \mathcal{P}^{\alpha_n}(\hat{\mathbf{\Omega}}_j)
\end{bmatrix}
\nonumber\\
\end{eqnarray}

The above equation is a realization of the sky-estimate with filter $\alpha = \alpha_1$. Similarly, application of several other filters, $\{\alpha\}$, would yield set of equations, which can be written in a matrix form and is the master equation for the present work.
\begin{equation}\label{eqn:master_conv}
\begin{bmatrix}
\langle S^{\alpha_1}(\hat{\mathbf{\Omega}}_i) \rangle\\
\langle S^{\alpha_2}(\hat{\mathbf{\Omega}}_i) \rangle\\
\langle S^{\alpha_3}(\hat{\mathbf{\Omega}}_i) \rangle\\
\cdot\\
\cdot\\
\cdot\\
\langle S^{\alpha_n}(\hat{\mathbf{\Omega}}_i) \rangle
\end{bmatrix}
= \begin{bmatrix}
\mathcal{C}^{\alpha_1 \alpha_1}(\hat{\mathbf{\Omega}}_i, \hat{\mathbf{\Omega}}_j) & \mathcal{C}^{\alpha_1 \alpha_2}(\hat{\mathbf{\Omega}}_i, \hat{\mathbf{\Omega}}_j) & \mathcal{C}^{\alpha_1 \alpha_3}(\hat{\mathbf{\Omega}}_i, \hat{\mathbf{\Omega}}_j) & \cdot \cdot \cdot & \mathcal{C}^{\alpha_1 \alpha_n}(\hat{\mathbf{\Omega}}_i, \hat{\mathbf{\Omega}}_j) \\
\mathcal{C}^{\alpha_2 \alpha_1}(\hat{\mathbf{\Omega}}_i, \hat{\mathbf{\Omega}}_j) & \mathcal{C}^{\alpha_2 \alpha_2}(\hat{\mathbf{\Omega}}_i, \hat{\mathbf{\Omega}}_j) & \mathcal{C}^{\alpha_2 \alpha_3}(\hat{\mathbf{\Omega}}_i, \hat{\mathbf{\Omega}}_j) & \cdot \cdot \cdot & \mathcal{C}^{\alpha_2 \alpha_n}(\hat{\mathbf{\Omega}}_i, \hat{\mathbf{\Omega}}_j) \\
\mathcal{C}^{\alpha_3 \alpha_1}(\hat{\mathbf{\Omega}}_i, \hat{\mathbf{\Omega}}_j) & \mathcal{C}^{\alpha_3 \alpha_2}(\hat{\mathbf{\Omega}}_i, \hat{\mathbf{\Omega}}_j) & \mathcal{C}^{\alpha_3 \alpha_3}(\hat{\mathbf{\Omega}}_i, \hat{\mathbf{\Omega}}_j) & \cdot \cdot \cdot & \mathcal{C}^{\alpha_3 \alpha_n}(\hat{\mathbf{\Omega}}_i, \hat{\mathbf{\Omega}}_j) \\
\cdot & \cdot & \cdot & \cdot \cdot \cdot & \cdot\\
\cdot & \cdot & \cdot & \cdot \cdot \cdot & \cdot\\
\cdot & \cdot & \cdot & \cdot \cdot \cdot & \cdot\\
\mathcal{C}^{\alpha_n \alpha_1}(\hat{\mathbf{\Omega}}_i, \hat{\mathbf{\Omega}}_j) & \mathcal{C}^{\alpha_n \alpha_2}(\hat{\mathbf{\Omega}}_i, \hat{\mathbf{\Omega}}_j) & \mathcal{C}^{\alpha_n \alpha_3}(\hat{\mathbf{\Omega}}_i, \hat{\mathbf{\Omega}}_j) & \cdot \cdot \cdot & \mathcal{C}^{\alpha_n \alpha_n}(\hat{\mathbf{\Omega}}_i, \hat{\mathbf{\Omega}}_j) \\
\end{bmatrix}
\begin{bmatrix}
\mathcal{P}^{\alpha_1}(\hat{\mathbf{\Omega}}_j)\\
\mathcal{P}^{\alpha_2}(\hat{\mathbf{\Omega}}_j)\\
\mathcal{P}^{\alpha_3}(\hat{\mathbf{\Omega}}_j)\\
\cdot\\
\cdot\\
\cdot\\
\mathcal{P}^{\alpha_n}(\hat{\mathbf{\Omega}}_j)\\
\end{bmatrix}
\end{equation}
The matrix formed by the assemblage of the estimates is obtained from equation (\ref{eqn:conv}), having dimensions ${(n\times \text{\# pixels})\times 1}$, forms the l.h.s of the master convolution equation (\ref{eqn:master_conv}). The square matrix with elements $\mathcal{C}^{\alpha \beta}(\hat{\mathbf{\Omega}}_i, \hat{\mathbf{\Omega}}_j)$, formulated below, in equation (\ref{eqn:coupling_matrix_elements}), is the \textit{coupling matrix}, $\mathcal{C}$. It couples the \textit{true} values of the angular power spectrum (signal from nature) with \textit{single-index} estimates and contains the information of the model spectrum, model sky-map and the detector noises. It has the dimension of ${(n\times \text{\# pixels})\times (n \times \text{\# pixels})}$, where $n$ is the total number of \textit{filters} used and $\text{\# pixels}$ refers to the resolution of the map (or no. of pixels). Here, $\{\alpha, \beta\} \in \{\alpha_1, \alpha_2, \alpha_3...\alpha_n\}$, denote the spectral indices and $\{i, j\}$ run through all the pixel values, pointing to different directions in the sky. Thus, the dimensions of the $\mathcal{C}$ matrix depends on the number of \textit{components} one is interested in probing.

\begin{equation}\label{eqn:coupling_matrix_elements}
\mathcal{C}^{\alpha \beta}(\hat{\mathbf{\Omega}}_i, \hat{\mathbf{\Omega}}_j) = \Lambda_{\alpha}(\hat{\mathbf{\Omega}}_i) \times \sum_{tf} \frac{H^{\alpha}(f) \ H^{\beta}(f)}{P_{1}(t;f) \ P_{2}(t;f)} \ \gamma^*(t,f;\hat{\mathbf{\Omega}}_i) \ \gamma(t,f;\hat{\mathbf{\Omega}}_j) 
\end{equation}

\begin{figure}[h!]
\begin{subfigure}[t]{.6\textwidth}
\centering
\includegraphics[width=\linewidth]{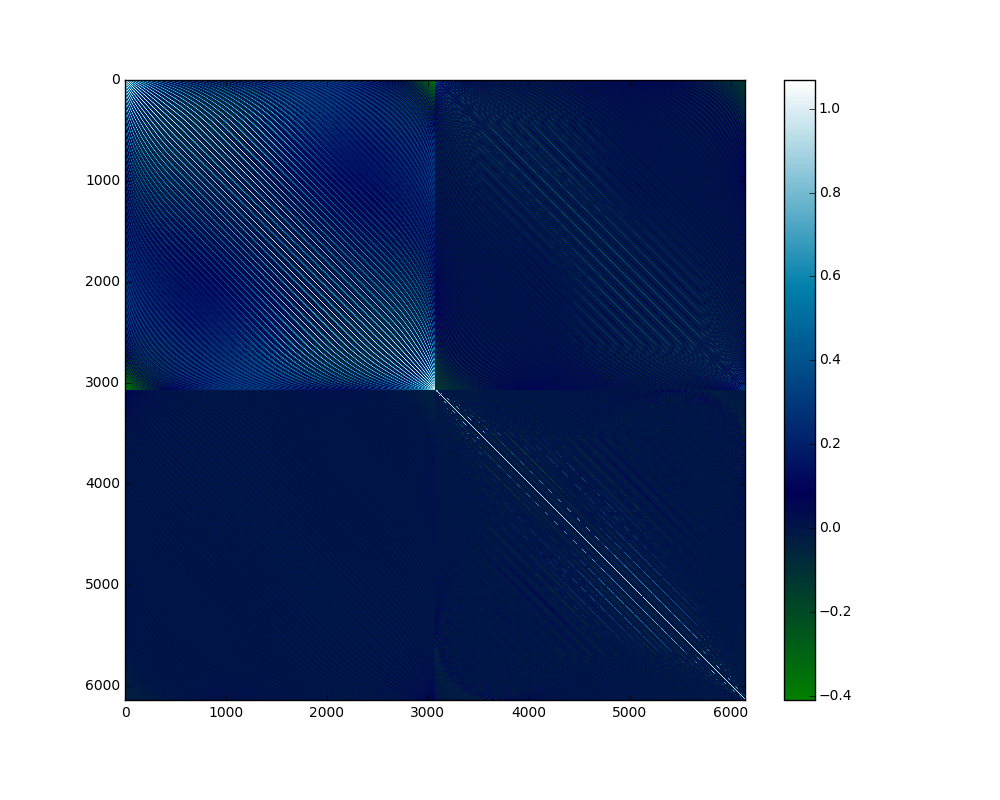}
\caption{For the design sensitivity of aLIGO.}\label{fig:Beam-design}
\end{subfigure}
\begin{subfigure}[t]{.6\textwidth}
\centering
\includegraphics[width=\linewidth]{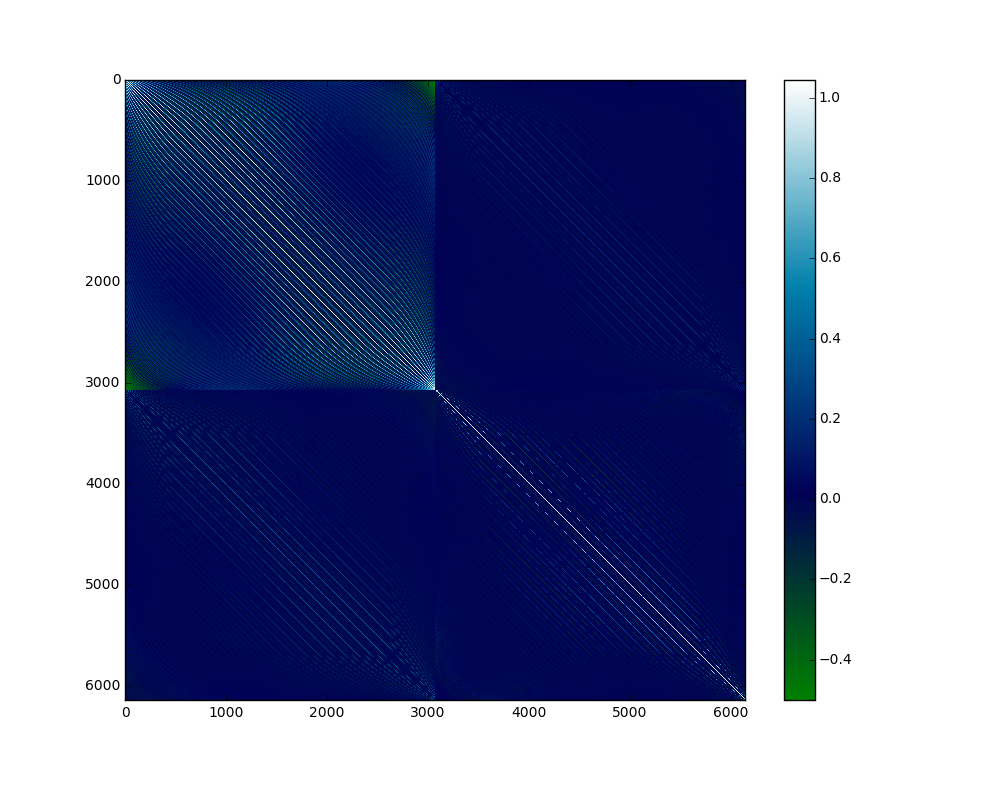}
\caption{For the O1-H1L1 data.}\label{fig:Beam-O1}
\end{subfigure}
\caption{Coupling matrix, $\mathcal{C}$, for a two component background $\alpha \in \{0, 3\}$.}
\end{figure}

Comparing with our previous work \cite{compSep}, this formula bears the similarity with the coupling matrix for an isotropic search. By the method of its construction, it is symmetric. Similarly, the overall error matrix ($\Sigma$), can be written as blocks of smaller $\sigma$ matrices,

\begin{equation}
\Sigma = 
\begin{bmatrix}
\sigma^{\alpha_1 \alpha_1}(\hat{\mathbf{\Omega}}_i, \hat{\mathbf{\Omega}}_j) & \sigma^{\alpha_1 \alpha_2}(\hat{\mathbf{\Omega}}_i, \hat{\mathbf{\Omega}}_j) & \sigma^{\alpha_1 \alpha_3}(\hat{\mathbf{\Omega}}_i, \hat{\mathbf{\Omega}}_j) & \cdot \cdot \cdot & \sigma^{\alpha_1 \alpha_n}(\hat{\mathbf{\Omega}}_i, \hat{\mathbf{\Omega}}_j) \\
\sigma^{\alpha_2 \alpha_1}(\hat{\mathbf{\Omega}}_i, \hat{\mathbf{\Omega}}_j) & \sigma^{\alpha_2 \alpha_2}(\hat{\mathbf{\Omega}}_i, \hat{\mathbf{\Omega}}_j) & \sigma^{\alpha_2 \alpha_3}(\hat{\mathbf{\Omega}}_i, \hat{\mathbf{\Omega}}_j) & \cdot \cdot \cdot & \sigma^{\alpha_2 \alpha_n}(\hat{\mathbf{\Omega}}_i, \hat{\mathbf{\Omega}}_j) \\
\sigma^{\alpha_3 \alpha_1}(\hat{\mathbf{\Omega}}_i, \hat{\mathbf{\Omega}}_j) & \sigma^{\alpha_3 \alpha_2}(\hat{\mathbf{\Omega}}_i, \hat{\mathbf{\Omega}}_j) & \sigma^{\alpha_3 \alpha_3}(\hat{\mathbf{\Omega}}_i, \hat{\mathbf{\Omega}}_j) & \cdot \cdot \cdot & \sigma^{\alpha_3 \alpha_n}(\hat{\mathbf{\Omega}}_i, \hat{\mathbf{\Omega}}_j) \\
\cdot & \cdot & \cdot & \cdot \cdot \cdot & \cdot\\
\cdot & \cdot & \cdot & \cdot \cdot \cdot & \cdot\\
\cdot & \cdot & \cdot & \cdot \cdot \cdot & \cdot\\
\sigma^{\alpha_n \alpha_1}(\hat{\mathbf{\Omega}}_i, \hat{\mathbf{\Omega}}_j) & \sigma^{\alpha_n \alpha_2}(\hat{\mathbf{\Omega}}_i, \hat{\mathbf{\Omega}}_j) & \sigma^{\alpha_n \alpha_3}(\hat{\mathbf{\Omega}}_i, \hat{\mathbf{\Omega}}_j) & \cdot \cdot \cdot & \sigma^{\alpha_n \alpha_n}(\hat{\mathbf{\Omega}}_i, \hat{\mathbf{\Omega}}_j) \\
\end{bmatrix},
\end{equation}
where the individual elements are given by the formula,
\begin{equation}
\sigma^{\alpha \beta}(\hat{\mathbf{\Omega}}_i, \hat{\mathbf{\Omega}}_j) = \frac{1}{4}\bigg(\frac{20 \pi^2}{3H_0^2}\bigg)^2 \ \bigg[\sum_{tf} \frac{H^{\alpha}(f) \ H^{\beta}(f)}{P_1(t;f) \ P_2(t;f)} \ \gamma^*(t,f;\hat{\mathbf{\Omega}}_i) \ \gamma(t,f;\hat{\mathbf{\Omega}}_j) \bigg]^{-1}. 
\end{equation}
For instance, for a two-component GW background with spectral index $\alpha = [0, 3]$ and pixel resolution of $\text{\# pixels} = 3072$, the \textit{coupling} matrix and the \textit{error} matrix is given below; both have dimensions $6144 \times 6144$. The $\mathcal{C}$ matrix, considering two \textit{components} for the case of advance-LIGO design sensitivity and O1-run is plotted as a 2-d surface plot in figure (\ref{fig:Beam-design}) and (\ref{fig:Beam-O1}) respectively.

\begin{equation}
\mathcal{C}=
\left[
\begin{array}{c|c}
\mathcal{C}^{(0 0)}(\hat{\mathbf{\Omega}}_i, \hat{\mathbf{\Omega}}_j) & \mathcal{C}^{(0 3)}(\hat{\mathbf{\Omega}}_i, \hat{\mathbf{\Omega}}_j) \\
\hline
\mathcal{C}^{(3 0)}(\hat{\mathbf{\Omega}}_i, \hat{\mathbf{\Omega}}_j) & \mathcal{C}^{(3 3)}(\hat{\mathbf{\Omega}}_i, \hat{\mathbf{\Omega}}_j)
\end{array}
\right]
\quad \text{and} \quad 
\Sigma = 
\left[
\begin{array}{c|c}
\sigma^{(0 0)}(\hat{\mathbf{\Omega}}_i, \hat{\mathbf{\Omega}}_j) & \sigma^{(0 3)}(\hat{\mathbf{\Omega}}_i, \hat{\mathbf{\Omega}}_j) \\
\hline
\sigma^{(3 0)}(\hat{\mathbf{\Omega}}_i, \hat{\mathbf{\Omega}}_j) & \sigma^{(3 3)}(\hat{\mathbf{\Omega}}_i, \hat{\mathbf{\Omega}}_j)
\end{array}
\right]
\end{equation}

\section{Case study and Results (sky maps)}\label{sec:results}
In this section we demonstrate our method, restricting to aLIGO design sensitivity and O1-data. We also inject mock values occasionally in the source map and CSD (or PSD) and detail out our results. For numerical computations, we assume a power-law background spectrum, $\mathcal{F}^{\alpha}(f)$, as discussed earlier in section (\ref{sub_sec:multiComponents}). The frequency range is chosen from a lower value of $f_l =$ 20 Hz to a maximum of $f_u =$ 500 Hz, with a frequency resolution $\Delta f =$ 2.0 Hz. We use folded data for our analysis \cite{folding}. It has the effect of compressing the entire observation data of several months to a single sidereal day. In other words, we have a total integration time of 86164 seconds. We remind the readers that here maps are represented with merely one dimensional arrays, where each array element refers to a particular direction in the sky as defined by the Healpix pixelization schemes. For our purpose of map making, we used the resolution or the pixel grid size ($\text{\# pixels}$) of 3072. We attempt deconvolution by using the built-in module, namely, the Sparse linear algebra, in the \textit{scipy} package of \textit{python}. It is a direct procedure for solving a linear system of equations. Notably, the \textit{conjugate gradient} method worked the best.

\subsection{Sky map recovery for a Single-index analysis}

\begin{figure}[h!]
\begin{subfigure}[t]{.55\textwidth}
\centering
\includegraphics[width=\linewidth]{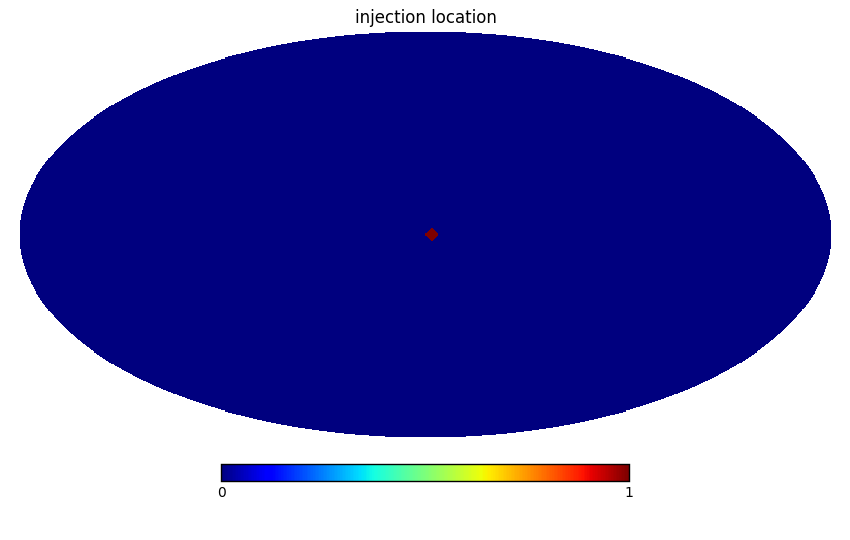}
        \caption{injection location}\label{fig:fig_a}
\end{subfigure}
\begin{subfigure}[t]{.55\textwidth}
\centering
\includegraphics[width=\linewidth]{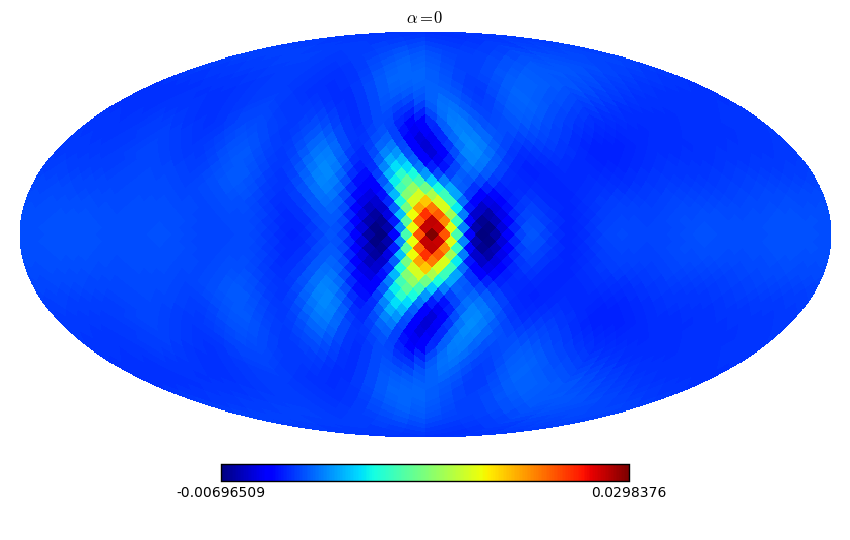}
\caption{$\alpha = 0$}\label{fig:fig_b}
\end{subfigure}
\begin{subfigure}[t]{.55\textwidth}
\centering
\includegraphics[width=\linewidth]{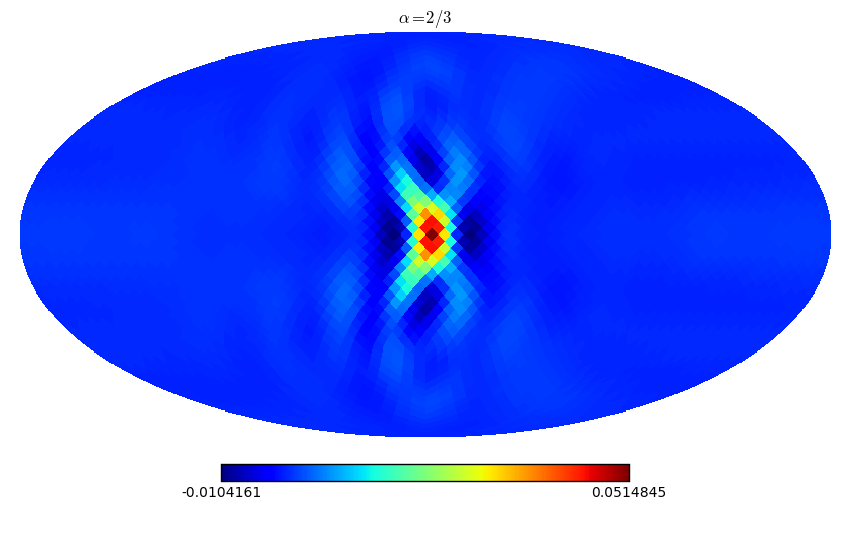}
        \caption{$\alpha = \frac{2}{3}$}\label{fig:fig_c}
\end{subfigure}
\begin{subfigure}[t]{.55\textwidth}
\centering
\includegraphics[width=\linewidth]{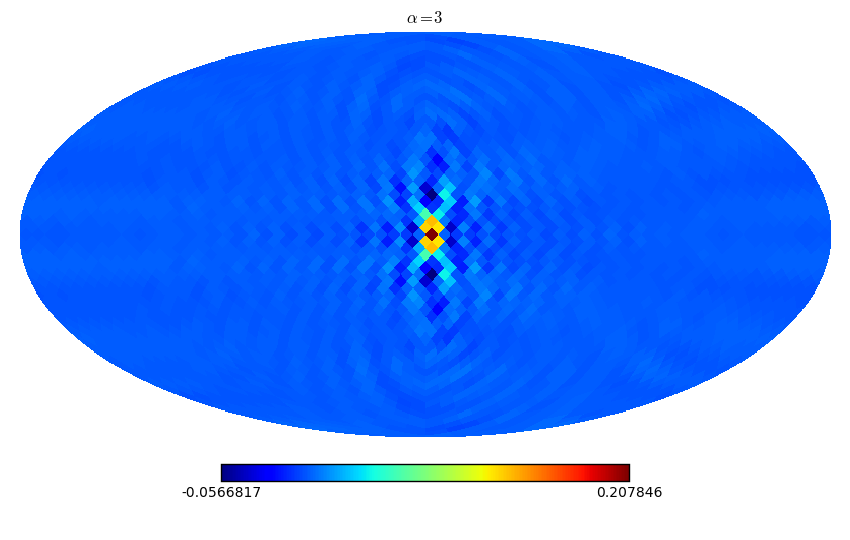}
        \caption{$\alpha = 3$}\label{fig:fig_d}
\end{subfigure}

\caption{Comparison of recovery amplitudes for a Single-index analysis: In (\subref{fig:fig_a}) we show the position of the injected point source each time this analysis is done with different filters and (\subref{fig:fig_b}), (\subref{fig:fig_c}) and (\subref{fig:fig_d}) represents the recovered (clean) sky-maps for various spectral index $\alpha \in \{0, 2/3, 3\}$. Notice difference in the amplitude of the recovered hot spot. We injected a hot spot at the center pixel of the sky-map for the purpose of simplified appearance.}\label{fig:comparison_Single_index}
\end{figure}


First, we demonstrate the general Radiometry analysis for a single point source and produce a sky-map for the same. This exercise is done for completeness and reflects a general fact that in the output (clean) map, refer figure (\ref{fig:comparison_Single_index}), the point source is recovered as a pattern of eight, with reduced amplitude compared to the injected ones, as expected for the LIGO baseline detectors. The exercise reveals the compatibility of our algorithm for a \textit{single-index} analysis. We perform with various spectral index, $\alpha = \{0, 2/3, 3\}$, one at a time, to emphasize this point. We choose the central pixel as a common injection location for all the three separate instances of the power-law spectrum. The analysis precisely involves inserting a \textit{single} bright source of certain amplitude (in this case, 1.0), in the CSD (or equivalently the PSD), at a location and then determining it using the algorithm developed. The algorithm works as expected and produces fair results for the case of single-component background.

\subsection{Comparison of both the schemes of recovery}
Here we profile the comparison plots between the two schemes viz., the existing \textit{single-index} search and our inquiry of \textit{joint-index} method.

\subsubsection{Using aLIGO O1-data}
First, we provide the actual result of the paper, showing a comparison study between the two schemes of recovery using the aLIGO O1-data, in figure (\ref{fig:comparison_fig_1}). We do not inject any source to the PSD for this study and recover the sky-maps for both the methods. We report no detection of SGWB and provide upper limits for the sky-maps regarding the power-law spectrum, $\{\mathcal{F}^0(f), \mathcal{F}^{2/3}(f), \mathcal{F}^3(f)\}$.

\subsubsection{Multiple injections with aLIGO design sensitivity}
Now, we present the results with injections of various brightness. We use the design sensitivity of the aLIGO and stick to the same power-law spectrum, but this time we study the effect of introducing some extended sources to the PSD through the equation (\ref{eqn:csd}). Their comparison result is presented in figure (\ref{fig:comparison_fig_2}). We inject three sources, (two extended sources, and a point source) viz., the dipole radiation, background noise from binaries in the Milky Way galaxy and a point source with the spectrum $\mathcal{F}^0(f), \mathcal{F}^{2/3}(f), \mathcal{F}^3(f)$ respectively. We then quantify the errors using the statistics Normalized Mean Square Error (NMSE), which is computed using the formula
\begin{equation}
\text{NMSE} = \frac{ ||\hat{\mathcal{P}} - \mathcal{P}||^2}{|| \mathcal{P}||^2},
\end{equation}
where $\mathcal{P}$ and $\hat{\mathcal{P}}$ are the injected (source) and estimated (clean) sky maps respectively. It should be noted that such a figure of merit is only possible when we have an injection in the source map. Otherwise, for the sky maps involving no injection, the errors are represented by a 2$\sigma$ upper limit as done for the O1 data in figure (\ref{fig:comparison_fig_1}). 
We provide the NMSE for the analysis, involving three injection to the source map in the table (\ref{tab:rms_error_with_injection}).

\begin{figure}[h!]
\begin{subfigure}[t]{.55\textwidth}
\centering
\includegraphics[width=\linewidth]{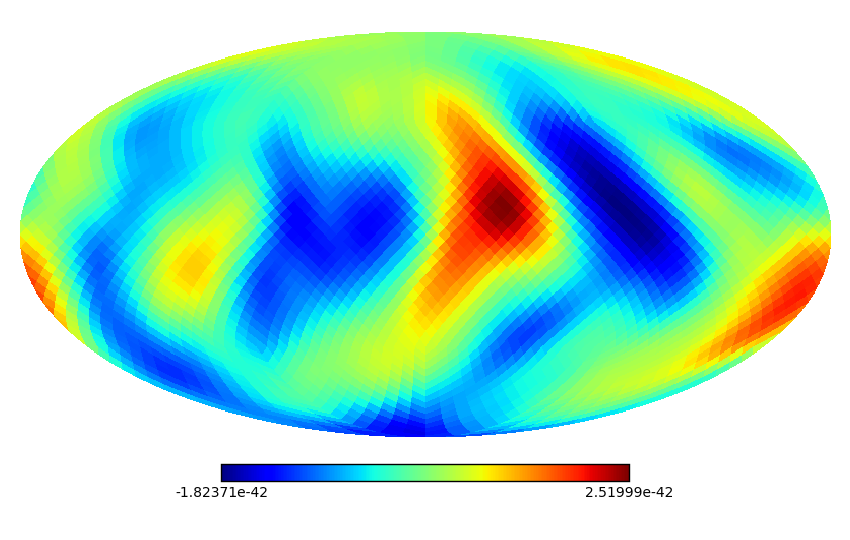}
\caption{Dirty map $\alpha = 0$}\label{fig:1}
\end{subfigure}
\begin{subfigure}[t]{.55\textwidth}
\centering
\includegraphics[width=\linewidth]{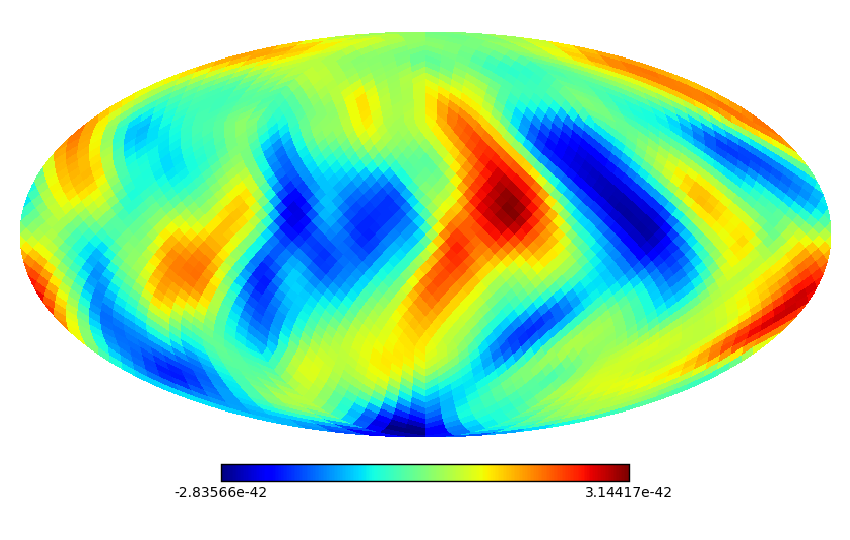}
\caption{Dirty map $\alpha = 2/3$}\label{fig:2}
\end{subfigure}
\begin{subfigure}[t]{.55\textwidth}
\centering
\includegraphics[width=\linewidth]{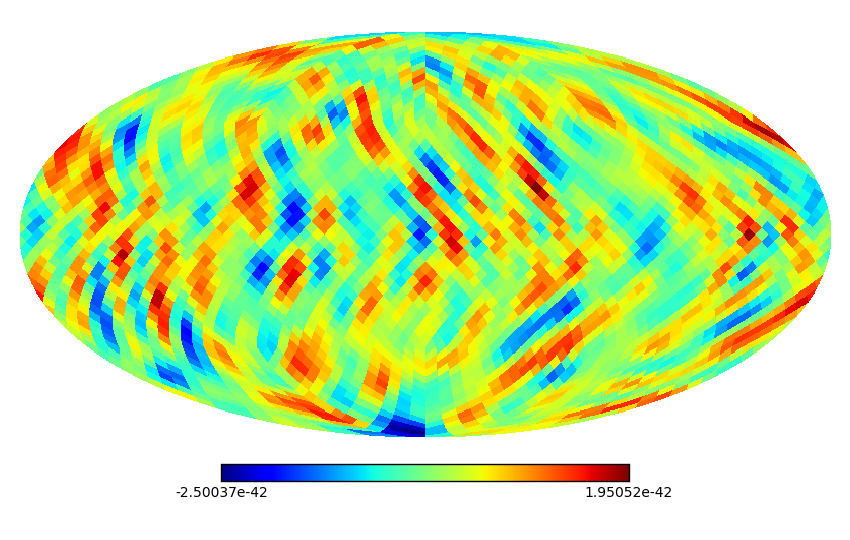}
\caption{Dirty map $\alpha = 3$}\label{fig:3}
\end{subfigure}
\caption{The multiple sub-plots illustrate the intermediate, raw maps (or commonly known as the Dirty map). The result was obtained working with the O1-H1L1 data, without any injection.}\label{fig:dirty_map_O1-H1L1}
\end{figure}

\begin{figure}[h!]
\begin{subfigure}[t]{.55\textwidth}
\centering
\includegraphics[width=\linewidth]{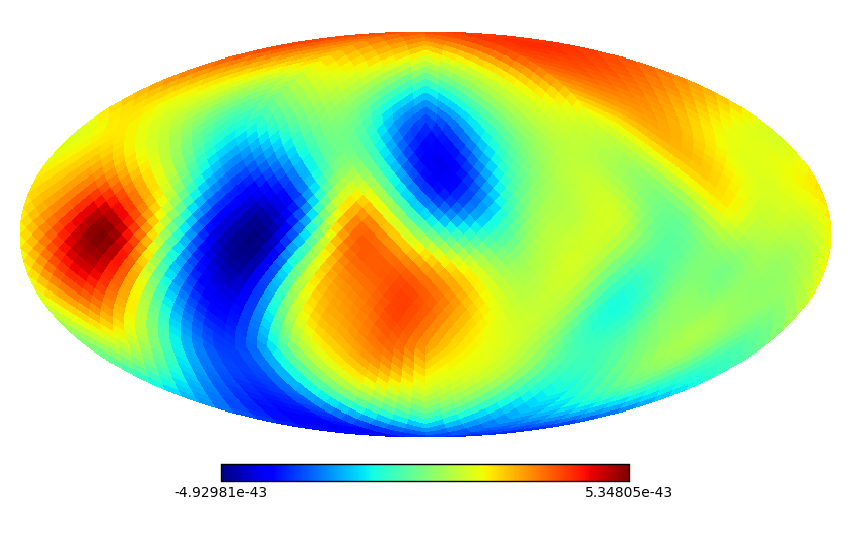}
\caption{Dirty map $\alpha = 0$}\label{fig:1}
\end{subfigure}
\begin{subfigure}[t]{.55\textwidth}
\centering
\includegraphics[width=\linewidth]{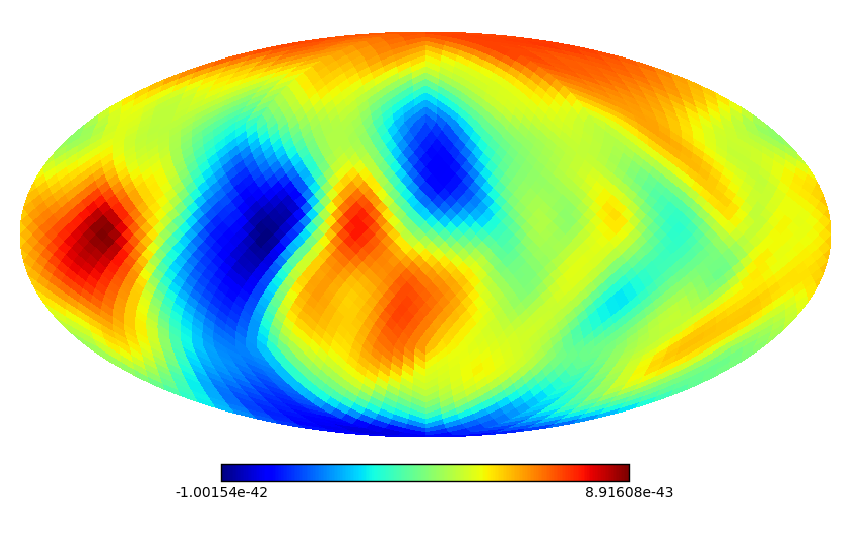}
\caption{Dirty map $\alpha = 2/3$}\label{fig:2}
\end{subfigure}
\begin{subfigure}[t]{.55\textwidth}
\centering
\includegraphics[width=\linewidth]{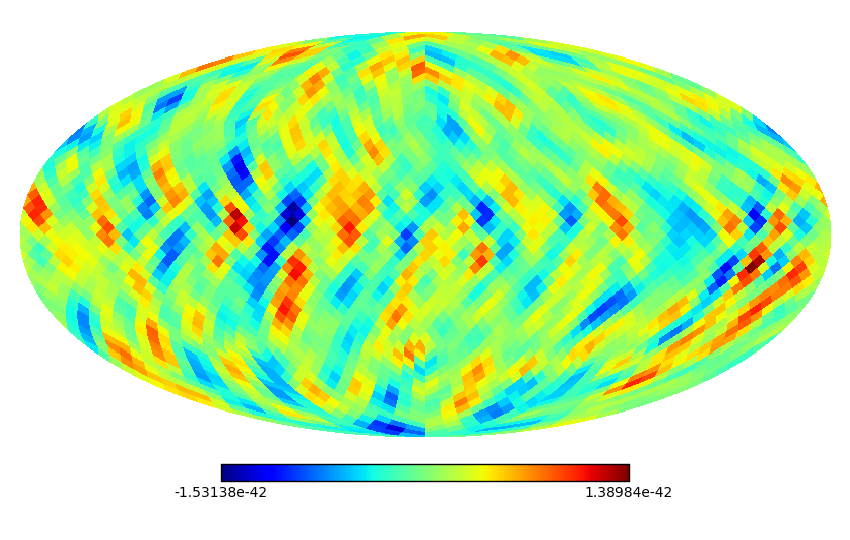}
\caption{Dirty map $\alpha = 3$}\label{fig:3}
\end{subfigure}
\caption{The multiple sub-plots illustrate the intermediate, raw maps (or commonly known as the Dirty map). The result was obtained working with the design sensitivity and noise only in the data, ie., without any injection.}\label{fig:dirty_map_O1-H1L1}
\end{figure}

\begin{figure}[h!]
%
%
%
\begin{subfigure}[t]{.55\textwidth}
\centering
\includegraphics[width=\linewidth]{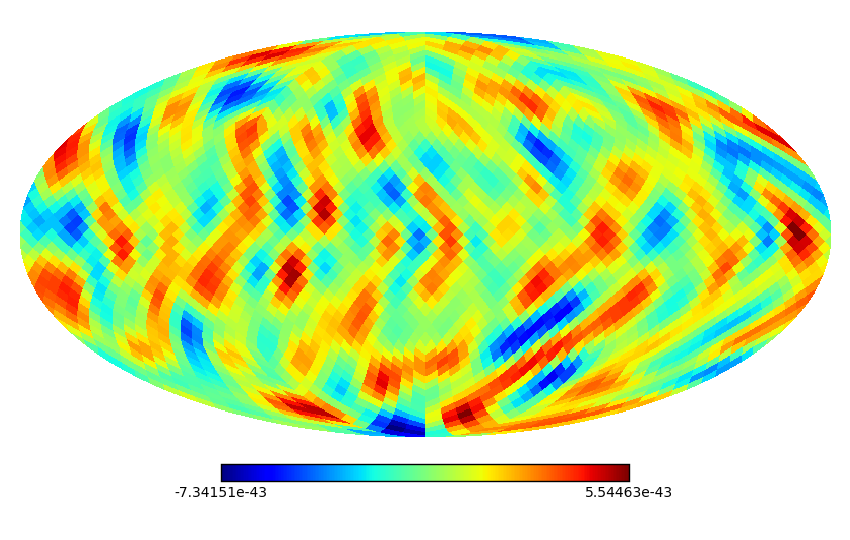}
\caption{Single index clean map $\alpha = 0$}\label{fig:3}
\end{subfigure}
\begin{subfigure}[t]{.55\textwidth}
\centering
\includegraphics[width=\linewidth]{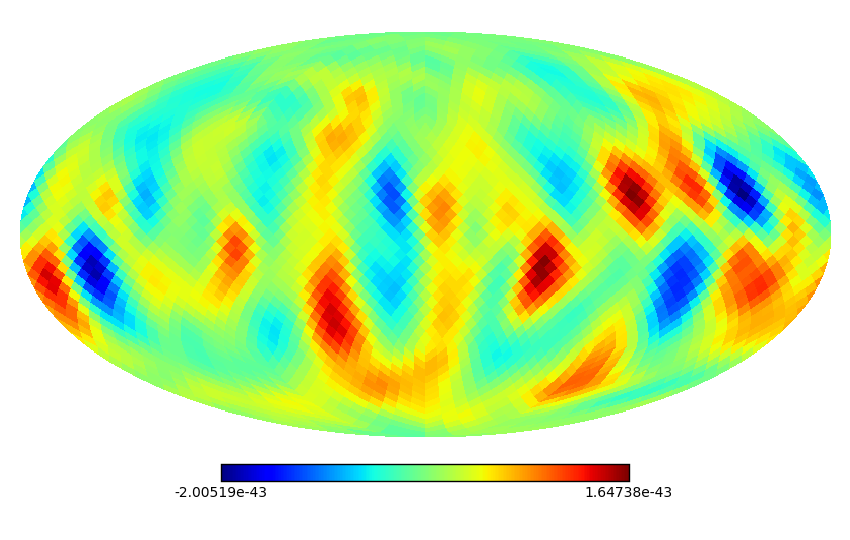}
\caption{Joint index clean map $\alpha = 0$}\label{fig:5}
\end{subfigure}
\begin{subfigure}[t]{.55\textwidth}
\centering
\includegraphics[width=\linewidth]{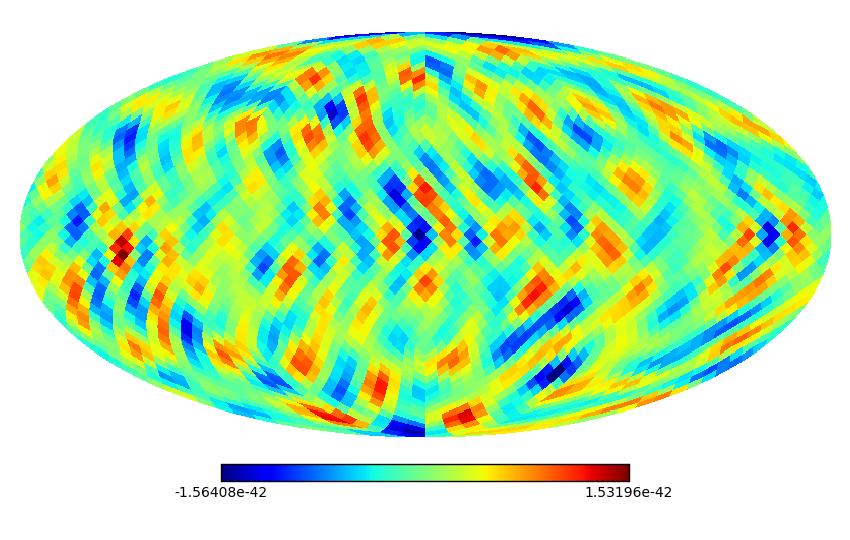}
\caption{Single index clean map $\alpha = 2/3$}\label{fig:3}
\end{subfigure}
\begin{subfigure}[t]{.55\textwidth}
\centering
\includegraphics[width=\linewidth]{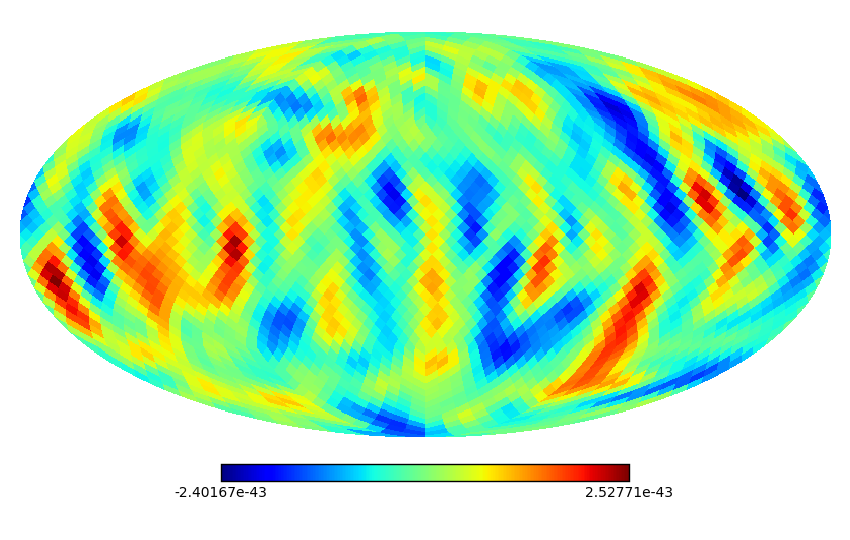}
\caption{Joint index clean map $\alpha = 2/3$}\label{fig:5}
\end{subfigure}
\begin{subfigure}[t]{.55\textwidth}
\centering
\includegraphics[width=\linewidth]{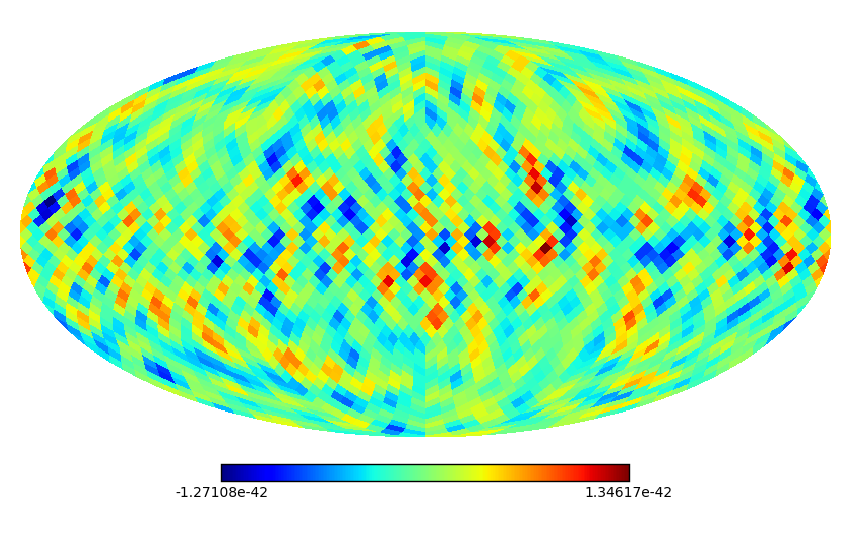}
\caption{Single index clean map $\alpha = 3$}\label{fig:4}
\end{subfigure}
\begin{subfigure}[t]{.55\textwidth}
\centering
\includegraphics[width=\linewidth]{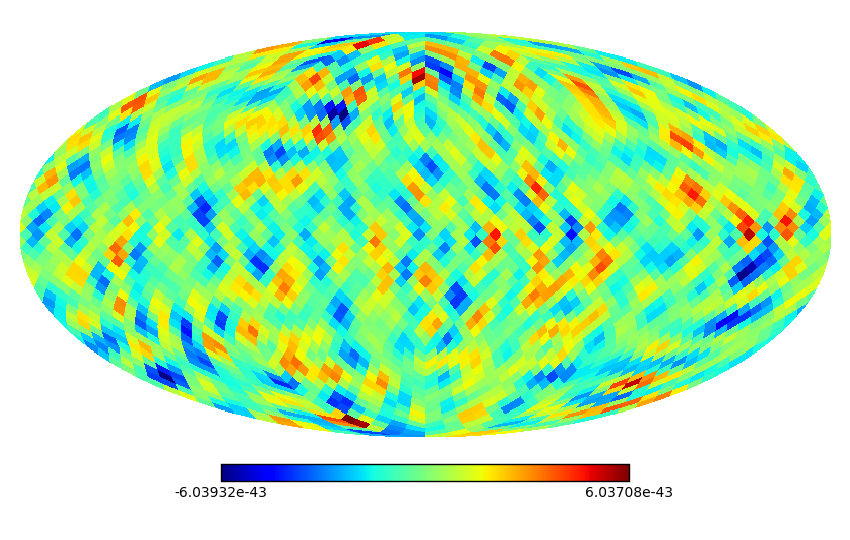}
\caption{Joint index clean map $\alpha = 3$}\label{fig:6}
\end{subfigure}
\caption{The multiple sub-plots contained in this figure compares the result from a Single-index (left panel) to that of a Joint-index (right panel) analysis described above. The corresponding dirty maps (intermediate results) are shown in figure (\ref{fig:dirty_map_O1-H1L1}). The analysis was done without any injection.}\label{fig:comparison_fig_1}
\end{figure}


\begin{figure}[h!]
\begin{subfigure}[t]{.55\textwidth}
\centering
\includegraphics[width=\linewidth]{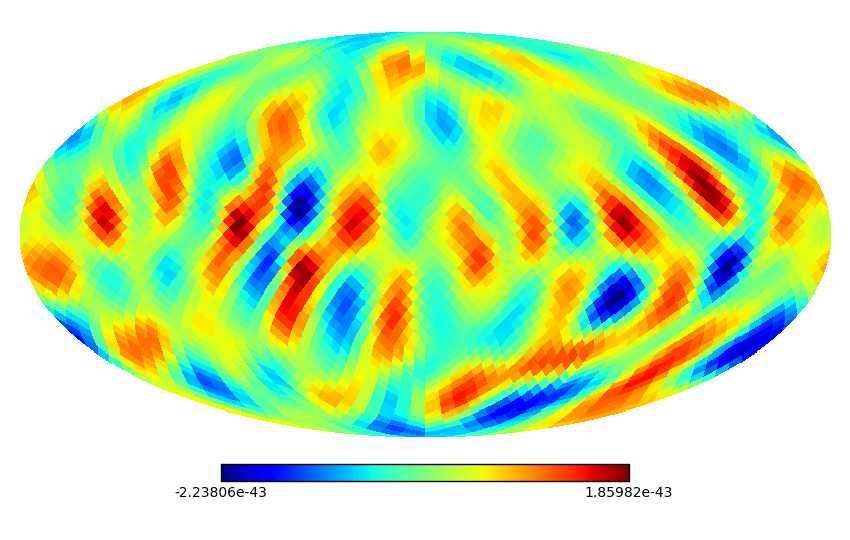}
\caption{Single index clean map $\alpha = 0$}\label{fig:3}
\end{subfigure}
\begin{subfigure}[t]{.55\textwidth}
\centering
\includegraphics[width=\linewidth]{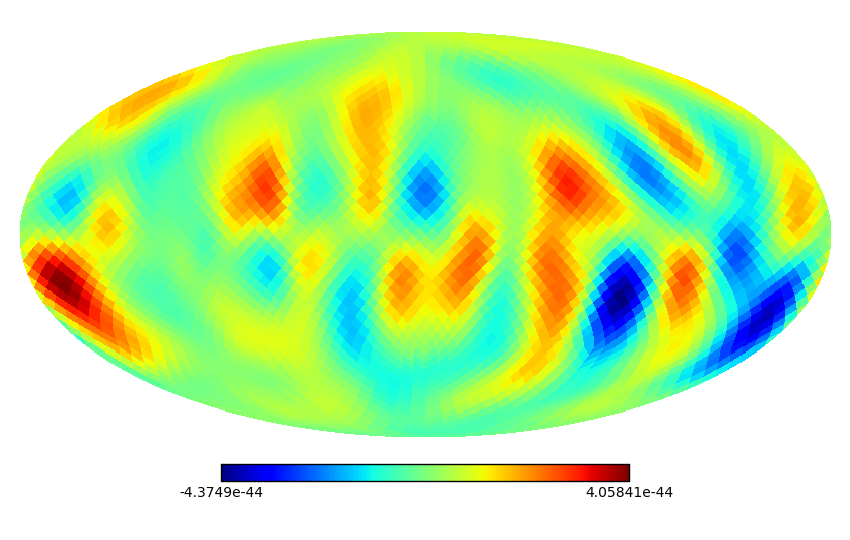}
\caption{Joint index clean map $\alpha = 0$}\label{fig:5}
\end{subfigure}
\begin{subfigure}[t]{.55\textwidth}
\centering
\includegraphics[width=\linewidth]{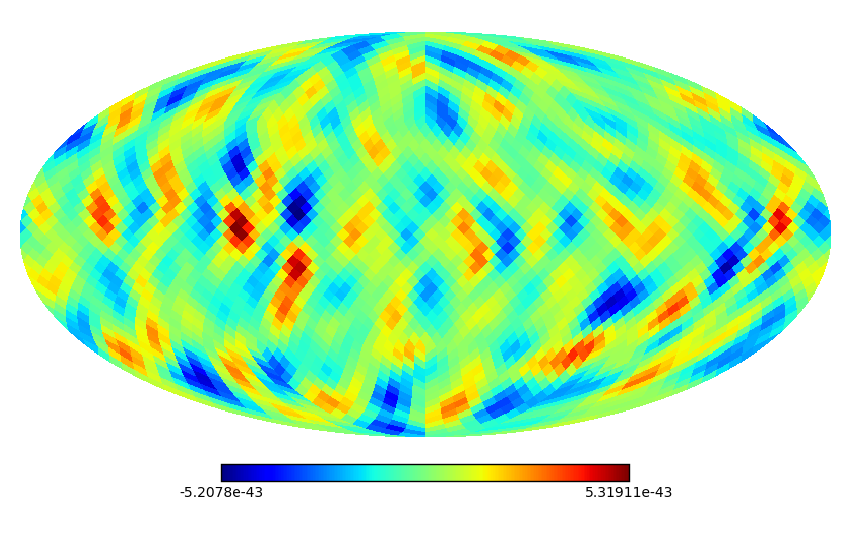}
\caption{Single index clean map $\alpha = 2/3$}\label{fig:3}
\end{subfigure}
\begin{subfigure}[t]{.55\textwidth}
\centering
\includegraphics[width=\linewidth]{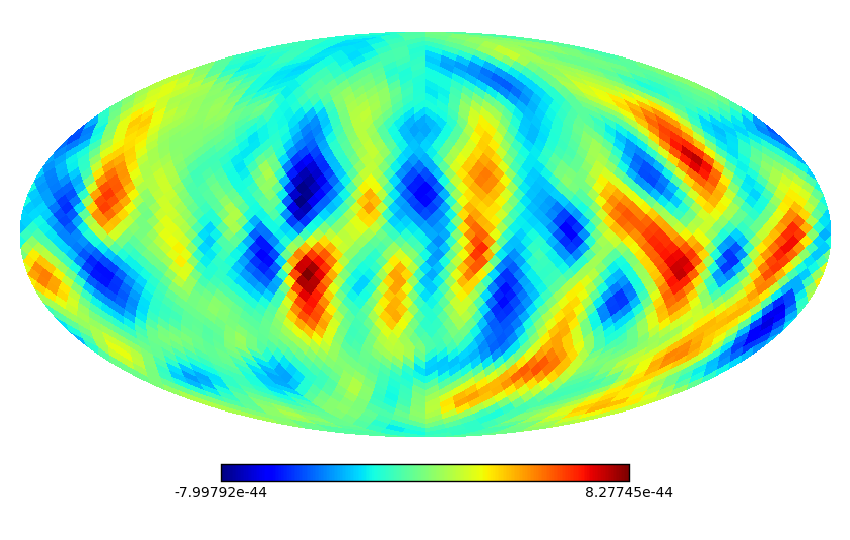}
\caption{Joint index clean map $\alpha = 2/3$}\label{fig:5}
\end{subfigure}
\begin{subfigure}[t]{.55\textwidth}
\centering
\includegraphics[width=\linewidth]{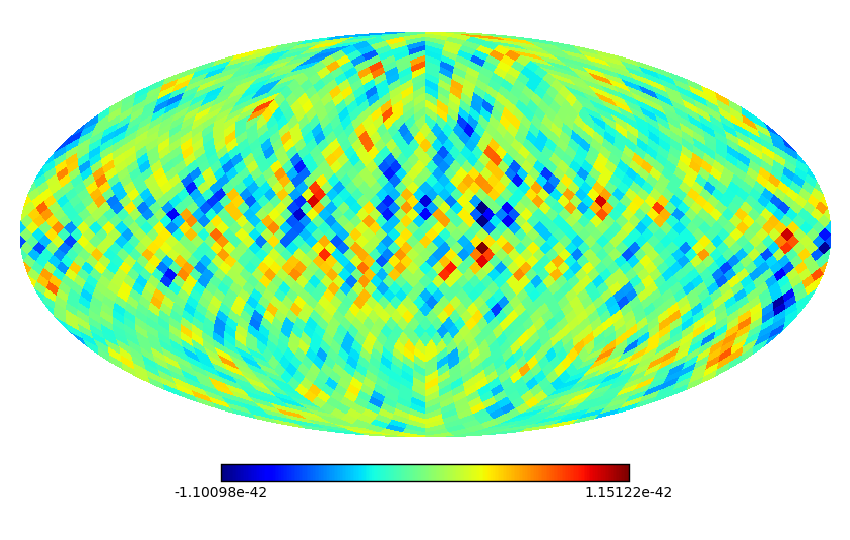}
\caption{Single index clean map $\alpha = 3$}\label{fig:4}
\end{subfigure}
\begin{subfigure}[t]{.55\textwidth}
\centering
\includegraphics[width=\linewidth]{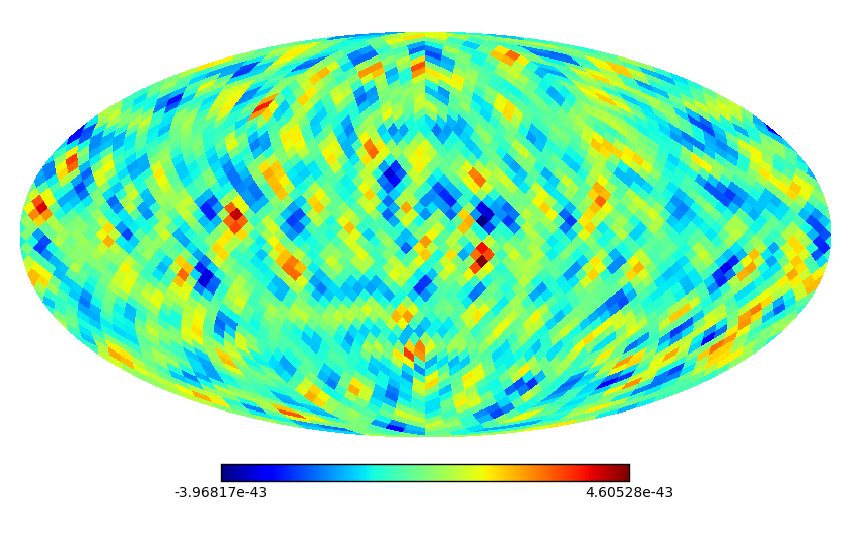}
\caption{Joint index clean map $\alpha = 3$}\label{fig:6}
\end{subfigure}
\caption{The multiple (noise only) sub-plots contained in this figure compares the result from a Single-index (left panel) to that of a Joint-index (right panel) analysis described above. The corresponding dirty maps (intermediate results) are shown in figure (\ref{fig:dirty_map_O1-H1L1}). The analysis was done without any injection.}\label{fig:comparison_fig_1}
\end{figure}

\begin{figure}[h!]
\begin{subfigure}[t]{.55\textwidth}
\centering
\includegraphics[width=\linewidth]{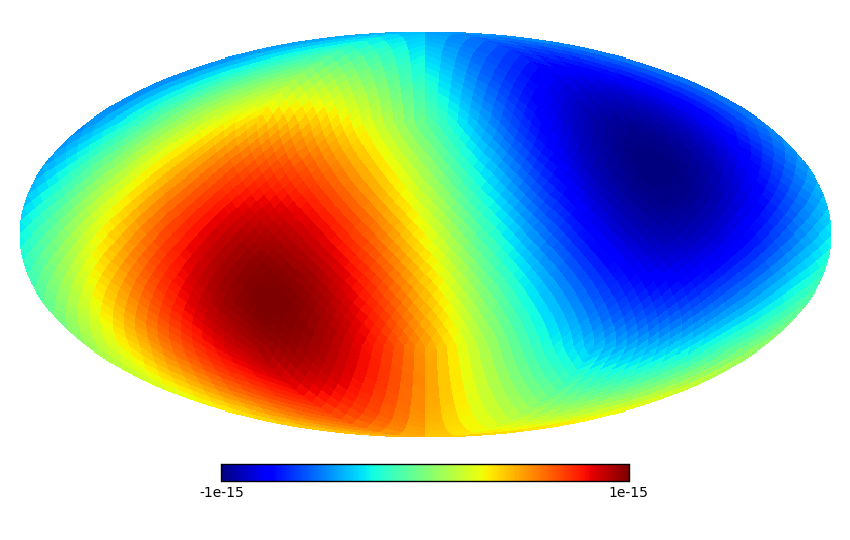}
\caption{Dipole source, $\alpha = 0$}
\end{subfigure}
\begin{subfigure}[t]{.55\textwidth}
\centering
\includegraphics[width=\linewidth]{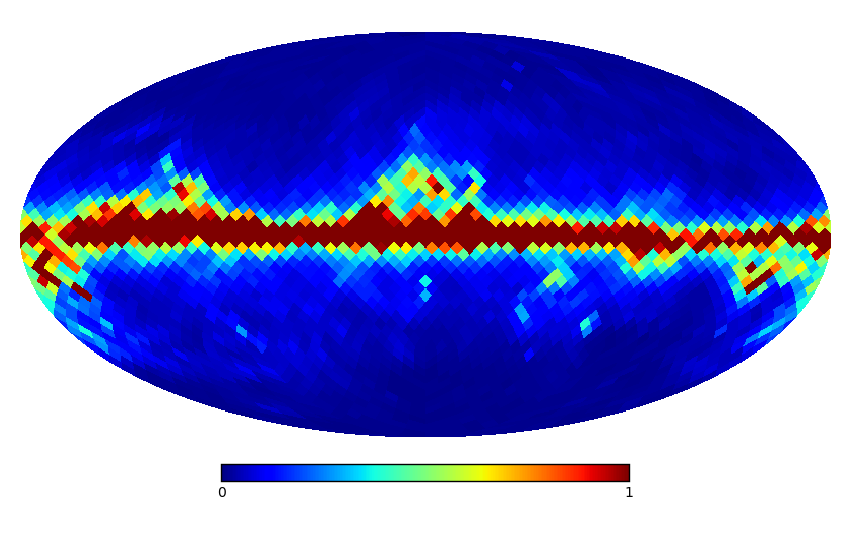}
\caption{Milky Way galaxy, $\alpha = 2/3$}
\end{subfigure}
\begin{subfigure}[t]{.55\textwidth}
\centering
\includegraphics[width=\linewidth]{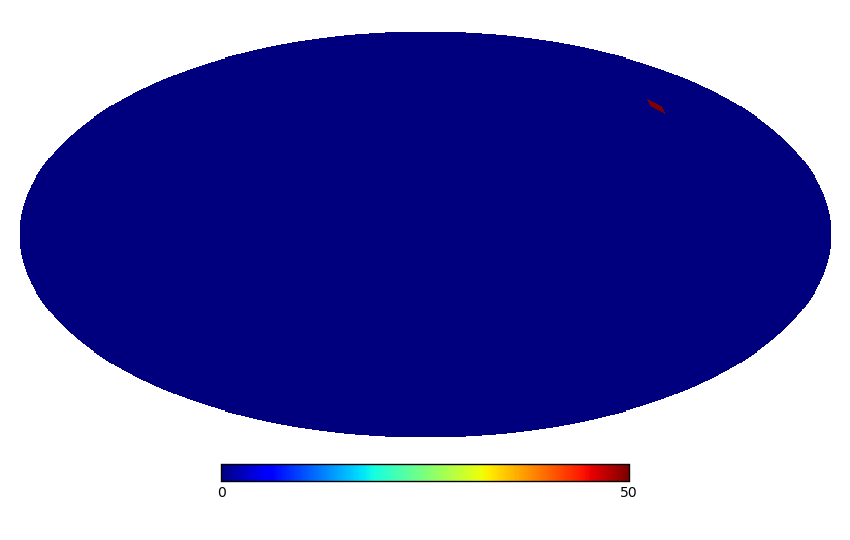}
\caption{A single point source, $\alpha = 3$}
\end{subfigure}
\begin{subfigure}[t]{.55\textwidth}
\centering
\includegraphics[width=\linewidth]{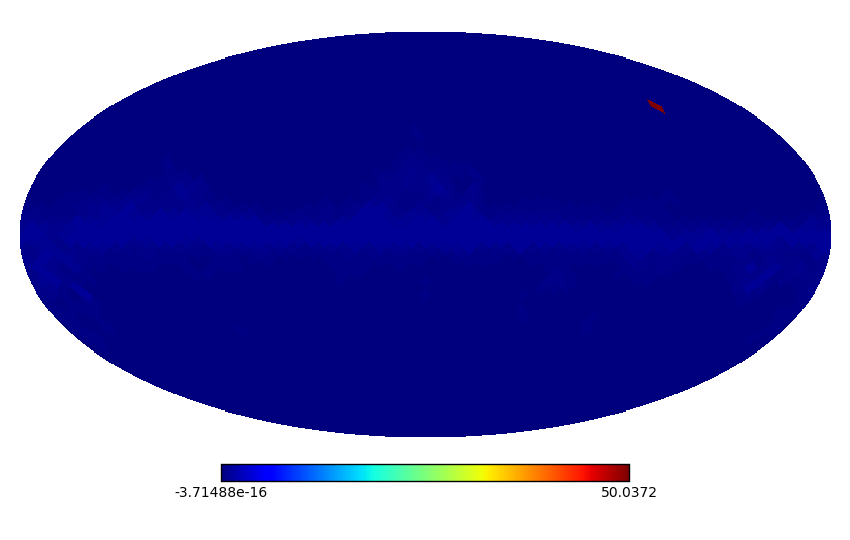}
\caption{Combined sources}
\end{subfigure}
\caption{Summary of various sources injected to the O1-H1L1 data stream.}\label{fig:source_map_O1-H1L1}
\end{figure}

\begin{figure}[h!]
\begin{subfigure}[t]{.55\textwidth}
\centering
\includegraphics[width=\linewidth]{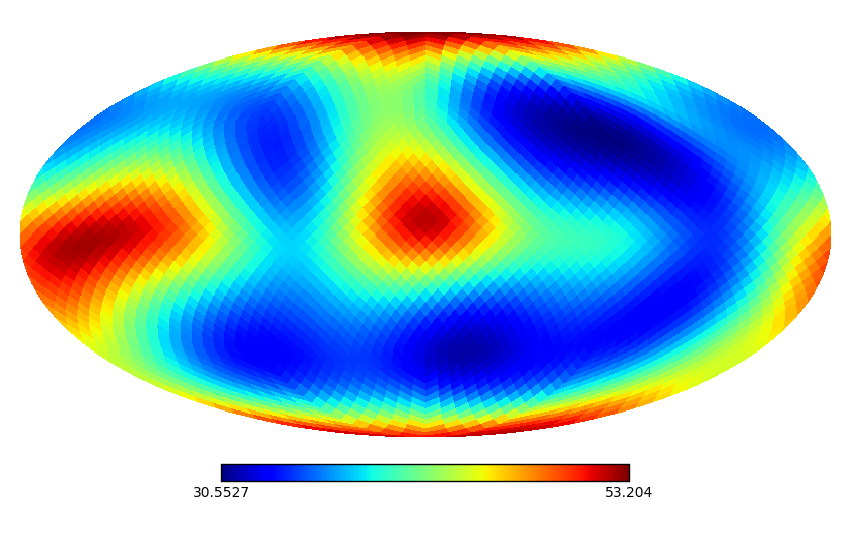}
\caption{Dirty map $\alpha = 0$}
\end{subfigure}
\begin{subfigure}[t]{.55\textwidth}
\centering
\includegraphics[width=\linewidth]{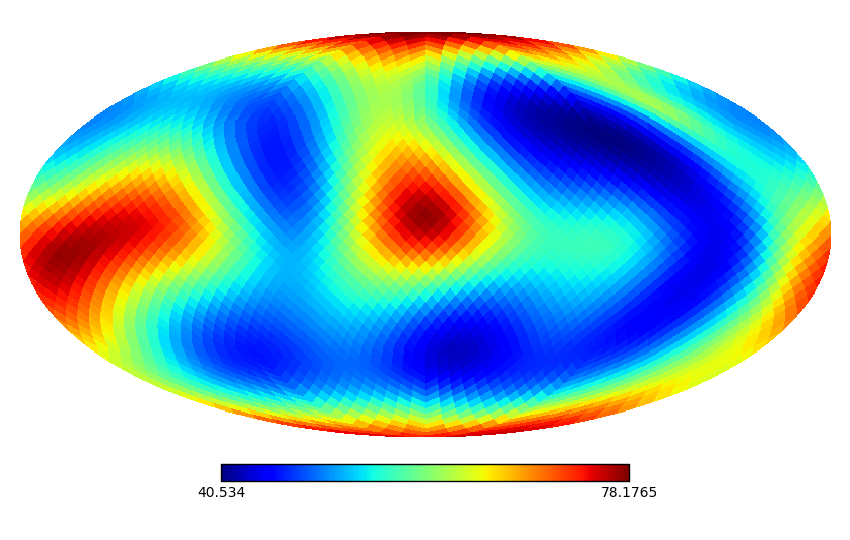}
\caption{Dirty map $\alpha = 0$}
\end{subfigure}
\begin{subfigure}[t]{.55\textwidth}
\centering
\includegraphics[width=\linewidth]{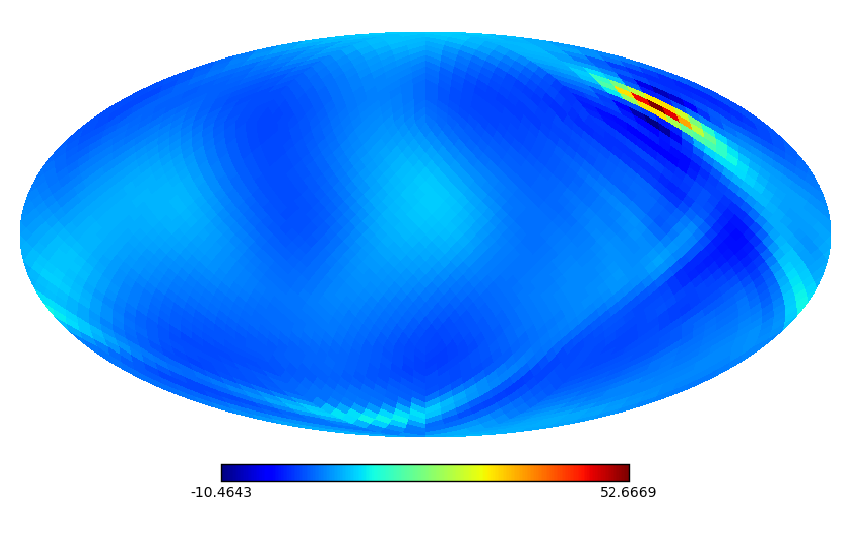}
\caption{Dirty map $\alpha = 3$}
\end{subfigure}
\caption{The sub-plots in this figure represent a similar intermediate results as the earlier figure (\ref{fig:dirty_map_O1-H1L1}). The only difference being that here we injected three sources (two extended and a point source) to the O1-H1L1 data using the equation (\ref{eqn:csd}), as shown in figure (\ref{fig:source_map_O1-H1L1}).}\label{fig:comparison_fig_2}
\end{figure}

\begin{figure}[h!]
\begin{subfigure}[t]{.55\textwidth}
\centering
\includegraphics[width=\linewidth]{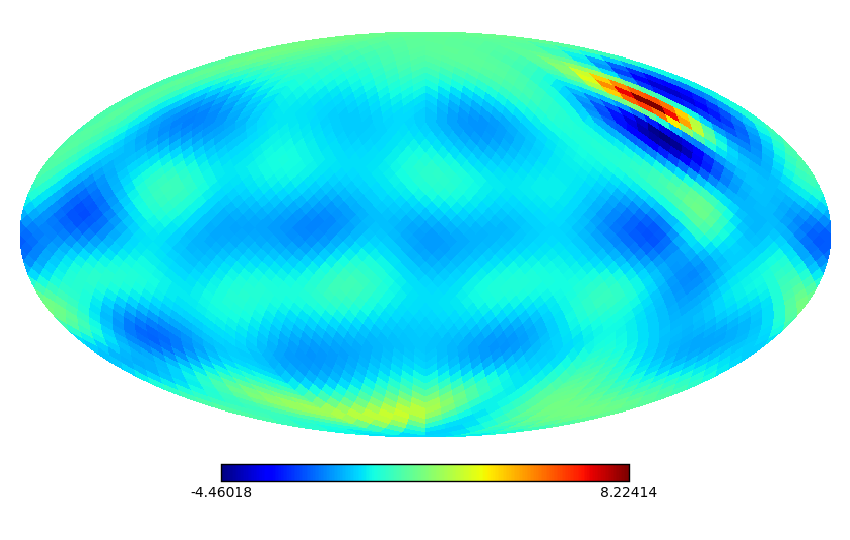}
\caption{Single index clean map $\alpha = 0$}
\end{subfigure}
\begin{subfigure}[t]{.55\textwidth}
\centering
\includegraphics[width=\linewidth]{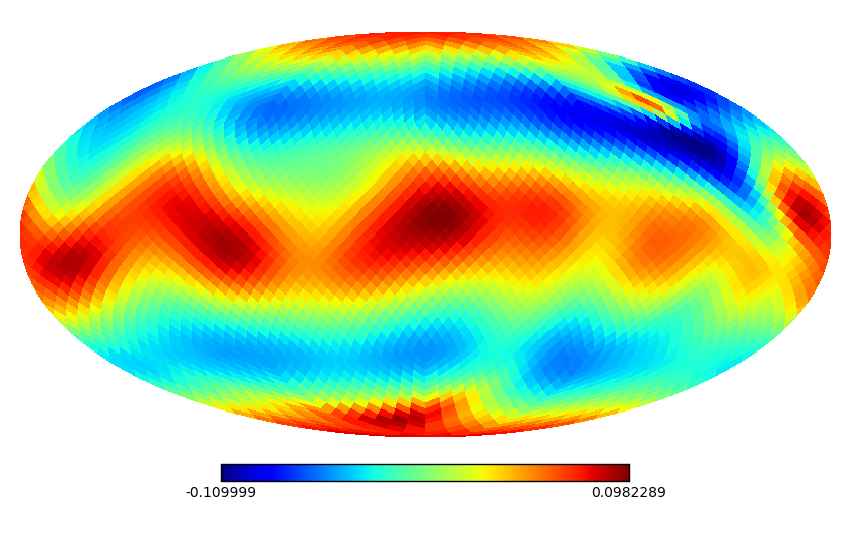}
\caption{Joint index clean map $\alpha = 0$}
\end{subfigure}
\begin{subfigure}[t]{.55\textwidth}
\centering
\includegraphics[width=\linewidth]{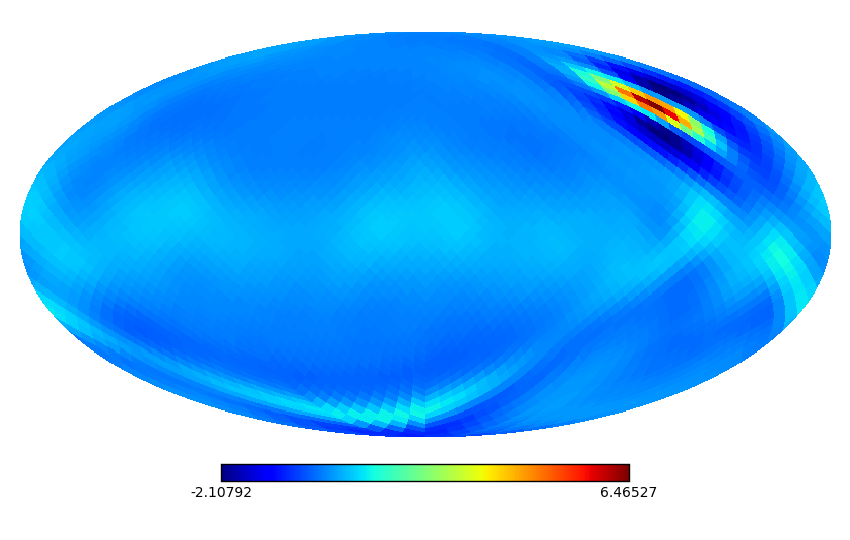}
\caption{Single index clean map $\alpha = 2/3$}
\end{subfigure}
\begin{subfigure}[t]{.55\textwidth}
\centering
\includegraphics[width=\linewidth]{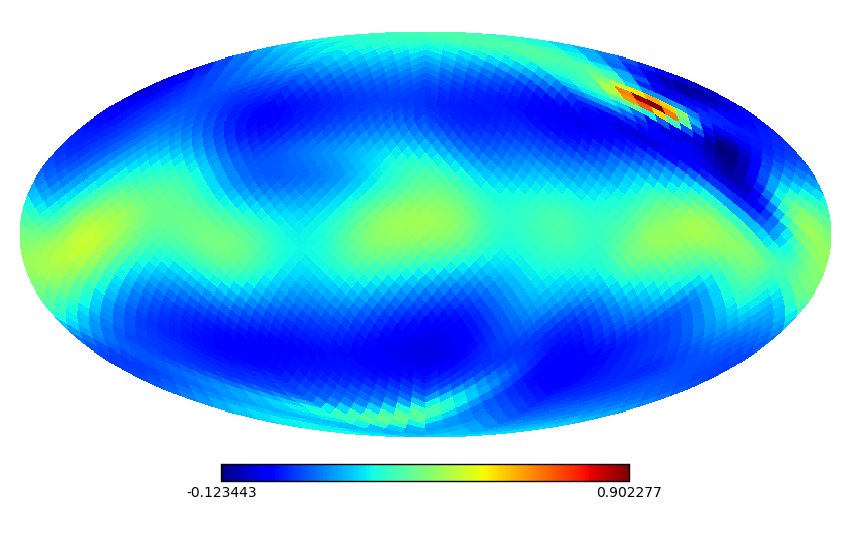}
\caption{Joint index clean map $\alpha = 2/3$}
\end{subfigure}
\begin{subfigure}[t]{.55\textwidth}
\centering
\includegraphics[width=\linewidth]{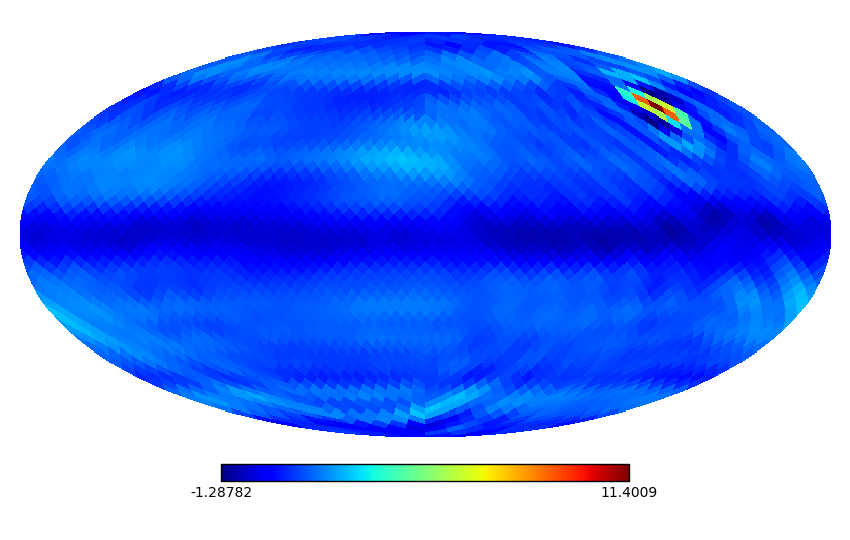}
\caption{Single index clean map $\alpha = 3$}
\end{subfigure}
\begin{subfigure}[t]{.55\textwidth}
\centering
\includegraphics[width=\linewidth]{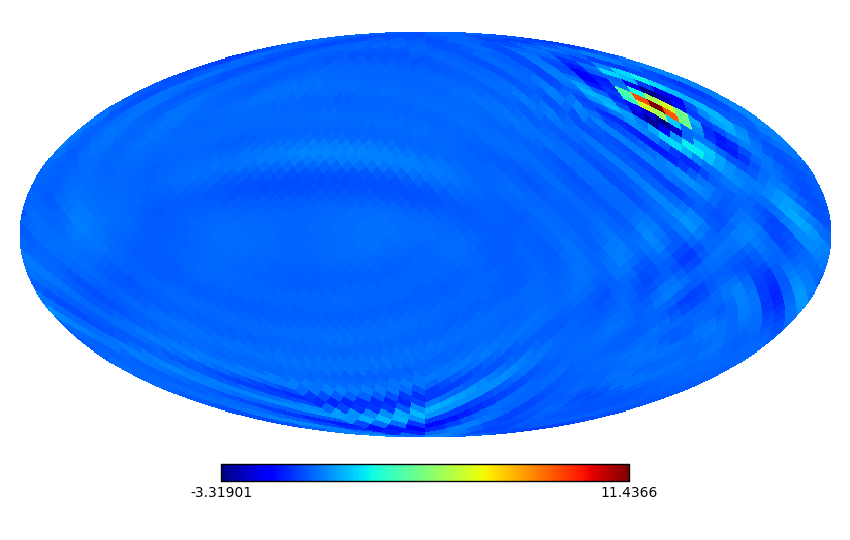}
\caption{Joint index clean map $\alpha = 3$}
\end{subfigure}
\caption{The sub-plots in this figure represent a similar result to figure (\ref{fig:comparison_fig_1}), viz., a comparison plot between the two schemes of investigation. In this instance we inject three sources, shown in figure (\ref{fig:source_map_O1-H1L1}), with power-law indices $\alpha = \{0, 2/3, 3\}$.}\label{fig:comparison_fig_2}
\end{figure}


\begin{table}[]
\centering
\begin{tabular}{|c|c|c|}
\hline
                  & \multicolumn{2}{c|}{\textbf{Normalized Mean Square Error}}           \\ \hline
\textbf{Spectrum} & \textbf{Single index analysis} &  \textbf{Joint multi-index analysis} \\ \hline
                $\mathcal{F}^0(f)$  & 2.804 $\times 10^{30}$                               & 6.325 $\times 10^{27}$                                     \\ \hline
                $\mathcal{F}^{2/3}(f)$  &  1.774                              &     0.401                                \\ \hline
                $\mathcal{F}^3(f)$  &  2.783                              &     0.840                                \\ \hline
\end{tabular}
\caption {The record shows the errors in the recovery process of clean maps produced in figure (\ref{fig:comparison_fig_2}). It demonstrates the comparison between the two scheme of investigation.}\label{tab:rms_error_with_injection}
\end{table}

\section{Conclusion}\label{sec:conclusion}
We have presented a method to separate anisotropies in different backgrounds using GW radiometry. Besides making use of the pixel basis, where contributions from each pixel are measured separately to generate the sky map, we also use LIGO O1 data folded to one sidereal day. Furthermore, at any given instance the pixels contain model dependent information as well as detector noise. For any real case scenarios, SGWB would be identified after analyzing and classifying out the individual resolved sources. In other words, pixels accommodating these clear-cut events need to be discarded entirely before beginning any study on SGWB. However for the case of unresolved GW events, one had to address the question, what is considered to be a proper detection? 

For numerical purposes, mainly the deconvolution routine is carried using a conjugate gradient algorithm, and the recovery of the injection demonstrates the validity of the algorithm.  Incorporating the component separation methods to PyStoch (very fast efficient method was introduced to probe the stochastic searches) pipeline as a module will improve all the analysis like the component separation method for directional searches discussed in this paper. By extending this work in that directions will reduce the computational cost and one can perform the entire component separation analysis more efficiently.

\acknowledgments
We thank the LIGO Scientific Collaboration for access to the data and gratefully acknowledge the support of the United States National Science Foundation (NSF) for the construction and operation of the LIGO Laboratory and Advanced LIGO as well as the Science and Technology Facilities Council (STFC) of the United Kingdom, and the Max-Planck-Society (MPS) for support of the construction of Advanced LIGO. Additional support for Advanced LIGO was provided by the Australian Research Council. We acknowledge the use of IUCAA LDG cluster Sarathi for the computational/numerical work. This research benefited from a grant awarded to IUCAA by the Navajbai Ratan Tata Trust (NRTT). S. M. acknowledges support from the Department of Science 
\& Technology (DST), India provided under the Swarna Jayanti Fellowships scheme.

\end{document}